\documentstyle{mn}
\input{psfig.sty}
\newcommand{\etal}{{et al.~}}
\newcommand{\lta}{\la}

\newcommand{\kmsmpc}{\>{\rm km}\,{\rm s}^{-1}\,{\rm Mpc}^{-1}}

\newcommand{\Mpc}{\>{\rm Mpc}}

\newcommand{\kpc}{\>{\rm kpc}}

\newcommand{\beq}{\begin{equation}}
\newcommand{\eeq}{\end{equation}}

\newcommand{\mpch}{\>h^{-1}{\rm {Mpc}}}

\newcommand{\rmd}{{\rm d}}

\newcommand{\msunh}{\>h^{-1}\rm M_\odot}

\newcommand{\apj}{ApJ}

\newcommand{\mnras}{MNRAS}

\newdimen\hssize
\newdimen\hdsize 
\begin{document}


\title[The Three-point Correlation Function of Galaxies]
      {The Three-point Correlation Function of Galaxies:
       Comparing Halo Occupation Models with Observations}
\author[Wang et al.]
       {Yu Wang$^{1,2}$, Xiaohu Yang$^{3,1,2}$, H.J. Mo$^{3}$,
        Frank C. van den Bosch$^{4}$, YaoQuan Chu$^{1,2}$
        \thanks{E-mail: wywa@mail.ustc.edu.cn}\\
      $^1$Center for Astrophysics, University of Science and Technology
          of China, Hefei, Anhui 230026, China\\
      $^2$National Astronomical Observatories, Chinese Academy of Science,
          Chao-Yang District, Beijing, 100012, P.R.China\\
      $^3$Department of Astronomy, University of Massachusetts,
          Amherst MA 01003-9305, USA\\
      $^4$Department of Physics, Swiss Federal Institute of
          Technology, ETH H\"onggerberg, CH-8093, Zurich, Switzerland}


\date{}


\maketitle

\label{firstpage}


\begin{abstract}
  We present models for the three-point correlation function (3PCF) of
  both dark matter and galaxies. We show that models based on the halo
  model  can  reasonably match  the  dark  matter  3PCF obtained  from
  high-resolution  $N$-body  simulations.    On  small  scales  ($r\la
  0.5\mpch$) the  3PCF is sensitive  to details regarding  the density
  distributions  of dark  matter  halos.  On  larger  scales ($r  \ga
  2.0\mpch$) the  results are very  sensitive to the abundance  of the
  few  most  prominent   halos.   Using  the  conditional  luminosity
  function, we also  construct models for the 3PCF  of galaxies, which
  we test against  large mock galaxy samples.  The  bias of the galaxy
  distribution with respect to the  dark matter, and the finite number
  of galaxies  that can be hosted by  individual halos, significantly
  reduce the normalized  three-point correlation function with respect
  to that of dark matter. Contrary to the 3PCF of the dark matter, the
  galaxy 3PCF is much less  sensitive to details regarding the spatial
  number density  distribution of galaxies in individual  halos or to
  the abundance  of the  few most massive  systems.  Finally,  we show
  that our  model based on  the conditional luminosity function  is in
  good agreement with results  obtained from the 2-degree Field Galaxy
  Redshift  Survey.  In  particular, the  model nicely  reproduces the
  observational  finding  that the  3PCF  for  early-type galaxies  is
  slightly   higher  than   that   of  late-type   galaxies, and  that
  there is no significant dependence of the 3PCF on galaxy luminosity.
\end{abstract}




\section{Introduction}
\label{sec:intro}

Understanding the formation and  evolution of large-scale structure in
the Universe is one of the most important goals in cosmology.  Most of
observationally  accessible information  comes to  us in  the  form of
(galaxy)  light,  and  large-scale  structure studies  have  therefore
predominantly  focused  on  analyzing  the  spatial  distribution  of
galaxies. The statistical tool that  is most commonly used to quantify
galaxy clustering are the  correlation functions (Peebles 1980). For a
Gaussian density field, the statistical properties are fully described
by the two-point correlation  function (or, equivalently, by the power
spectrum in Fourier space), and all reduced higher order correlations 
are zero.   However,    even though  the initial perturbations  in the
density field out of which  all structure formed are generally thought
to  be  Gaussian, the  present-day  distribution  of  galaxies in  the
Universe is  expected to be  non-Gaussian.  There are two  reasons for
this.   First of  all, typical  density perturbations  on  scales $\la
10\mpch$ have already become non-linear  at the present time. On these
scales, the non-linear dynamics  can create non-Gaussian fluctuations. 
Secondly, galaxies may  be biased with respect to  the underlying mass
distribution,  which  may produce  additional  non-Gaussian features.  
Therefore, there  has been much interest in  studying the higher-order
correlation   functions  of   galaxies,  especially   the  three-point
correlation function (3PCF)  or equivalently, its Fourier counterpart,
the  bi-spectrum (Peebles  \& Groth  1975;  Peebles 1980;  Jing, Mo  \&
B\"orner  1991;  Jing  \&   B\"orner  1997;  Scoccimarro  \etal  1998;
Buchalter  \& Kamionkowski  1999; Bernardeau  \etal 2002;  Verde \etal
2002).

Theoretically, the 3PCF  of the mass distribution in  the Universe has
been considered  by various authors using either  analytical models or
$N$-body simulations (e.g.  Fry 1984;  Matsubara \& Suto 1994; Suto \&
Matsubara  1994;  Jing  \&   B\"orner  1997,  1998;  Gazta\~{n}aga  \&
Bernardeau   1998;   Frieman  \&   Gazta\~{n}aga   1999;  Barriga   \&
Gazta\~{n}aga 2002;  Takada \& Jain 2003). Theoretical  models for the
3PCF of galaxies, however, are  much more difficult to obtain, as they
depend on  the details about how  galaxies form in  the cosmic density
field.   The simplest  model,  in  which galaxies  are  assumed to  be
linearly   biased  with  respect   to  the   mass,  is   obviously  an
over-simplification.

According to current the cold  dark matter (CDM) scenario of structure
formation, most of the mass in  the Universe is expected to be in dark
matter  halos. These  are  quasi-equilibrium systems  of dark  matter
particles,  formed through  non-linear gravitational  collapse.  Since
accurate analytical  models are now  available for the  mass function,
spatial clustering, and  density profile of the halo  population (e.g. 
Mo \&  White 2002  and references  therein), there has  been a  lot of
effort in recent years in constructing the so-called halo model, which
describes the  dark matter  mass distribution solely  in terms  of its
dark matter  building blocks (Ma \&  Fry 2000; White  2001; Kang \etal
2002; Mo, Jing  \& B\"orner 1997; Cooray \&  Sheth 2002 and references
therein).  In  the CDM cosmogony, galaxies and  other luminous objects
are assumed to form by  cooling and condensation of the baryons within
halos (White \& Rees 1978).   Hence, the distribution of galaxies can
be linked  to the halo model  if it is  combined with a model  for the
formation  of  galaxies  in  individual  halos.   Unfortunately,  the
physics of galaxy  formation are still poorly understood.   One way to
make progress, without a detailed  theory for how galaxies form, is to
model  the statistics  of  galaxy occupation  numbers  in dark  matter
halos.   Many recent  investigations have  used such  halo occupation
models  to study  various aspects  of galaxy  clustering (Jing,  Mo \&
B\"orner 1998;  Peacock \& Smith 2000; Seljak  2000; Scoccimarro \etal
2001; White  2001; Jing, B\"orner  \& Suto 2002; Bullock,  Wechsler \&
Somerville 2002;  Berlind \& Weinberg 2002; Scranton  2002; Kang \etal
2002; Marinoni \& Hudson 2002;  Kochanek \etal 2003; Zheng \etal 2002;
Berlind \etal 2003; Magliocchetti \& Porciani 2003; Yan \etal 2003).

In  this  paper, we  investigate  the 3PCF  of  both  dark matter  and
galaxies using the halo  model combined with the so-called conditional
luminosity function (hereafter CLF).  The CLF formalism was introduced
by Yang \etal (2003) and van den Bosch \etal (2003) as an extension of
the typical halo occupation  models.  It not only contains information
about the number of galaxies  per halo, but also on their luminosities
and  morphological  types. Using  the  CLF,  therefore,  allows us  to
investigate  how the  3PCF  of  galaxies depends  on  galaxy type  and
luminosity.  Our purpose  is twofold.   First, we  use high-resolution
$N$-body  simulations and  realistic mock  galaxy  samples constructed
from  them to  check the  accuracy of  the model  predictions  for the
3PCFs.  Secondly, we construct  detailed mock galaxy redshift surveys,
and compare the resulting 3PCFs  with those obtained from the 2 degree
Field Galaxy  Redshift Survey  (2dFGRS, Colless \etal  2003).  Earlier
investigations, based on simple assumptions about the relation between
galaxies  and dark  matter,  showed  that the  3PCF  predicted by  the
current  `concordance'  cosmology  is  significantly higher  than  the
observations indicate  (Jing \& B\"orner  1998, 2004).  We show  that, 
using  our   more     realistic    mock samples  based on the CLF, the 
discrepancy between model and observation can be alleviated.

This  paper is  organized as  follows.  In  Section  \ref{sec:HOD}, we
outline  the halo  occupation distribution  (HOD) models.   We present
halo-based models  for the two- and  three-point correlation functions
in  Sections  3  and   4,  respectively.   Comparisons  between  model
predictions   and    simulation   results   are    made   in   Section
\ref{sec:results}.   In   Section  \ref{sec:obs},  we   compare  model
predictions  of the  redshift-space 3PCFs  obtained from  mock samples
with the observational results  obtained from the 2dFGRS.  Finally, we
draw conclusions in Section \ref{sec:concl}.

\section{Overview of the halo model}
\label{sec:HOD}

The  basic  idea  of  the  halo  model is  to  describe  the  evolved,
non-linear dark matter distribution  in terms of halos with different
masses.  On strongly non-linear  scales, the clustering of dark matter
can be understood  in terms of the actual  density profiles of halos,
while on larger,  linear scales, it can be understood  in terms of the
spatial distribution of  dark matter halos (see Cooray  \& Sheth 2002
and  references therein).   The  halo model  contains three  essential
ingredients, which we review below.

\subsection{Halo mass function}

The mass function of dark matter halos, $n(M){\rm d}M$, describes the
number density of  dark matter halos as a function  of halo mass. The
Press-Schechter  formalism   (Press  \&  Schechter   1974)  yields  an
analytical estimate for $n(M)$, and we use the form given in Sheth, Mo
\& Tormen (2001):
\begin{equation}
\label{halomf} 
n(M) \, {\rm d}M = {{\overline \rho} \over M^2} \nu
f(\nu) \, \left| {{\rm d} {\rm ln} \sigma \over {\rm d} {\rm ln}
M}\right| {\rm d}M,
\end{equation}
where $\bar{\rho}$ is the mean  matter density of the Universe, $\nu =
\delta_{\rm c}/\sigma(M)$,  and  $\delta_{\rm c}$  is  the  critical  
over-density
required for collapse.  The quantity $\sigma(M)$ in the above equation
is the linear rms mass fluctuation on mass scale $M$ and $f(\nu)$ is a
function of $\nu$:
\begin{equation}
\label{fnuST} \nu \, f(\nu) = 2 A \,\left(1 + {1\over
\nu'{^{2q}}}\right)\ \left({\nu'{^2}\over 2\pi}\right)^{1/2}
\exp\left(-{\nu'{^2} \over 2}\right)\,
\end{equation}
with  $\nu'=\sqrt{a}\,\nu$, $a=0.707$, $q=0.3$  and $A\approx  0.322$. 
The  resulting  mass  function  has  been shown  to  be  in  excellent
agreement  with numerical  simulations,  as long  as  halo masses  are
defined as the masses inside  a sphere with an average over-density of
about  $180$ (Sheth  \& Tormen  1999;  Jenkins \etal  2001).  In  what
follows, we define the radius of  this sphere as $r_{\rm 180}$ and the
corresponding volume as $V_{\rm 180}$.

\subsection{Halo density profile}

The  dark  matter density  profile,  $\rho  (r)$,  describes the  mass
distribution  within  individual dark  matter  halos (e.g.,  Navarro,
Frenk  \& White  1997, hereafter  NFW;  Moore \etal  1998; Jing  2002;
Bullock \etal 2001).  We assume that $\rho(r)$ has the NFW form
\begin{equation}
\label{NFW} 
\rho(r) = \frac{{\overline \delta} {\overline \rho}}
{(r/r_{\rm s})(1+r/r_{\rm
 s})^{2}},
\end{equation}
where $r_{\rm s}$ is a characteristic radius, and $\bar{\delta}$ is  a
dimensionless amplitude  which can be  expressed in terms of  the halo
concentration parameter $c=r_{\rm 180}/r_{\rm s}$ as
\begin{equation}
\label{overdensity} \bar{\delta} = {180 \over 3} \, {c^{3} \over
{\rm ln}(1+c) - c/(1+c)}.
\end{equation}
Numerical simulations have shown that $c$ is correlated with halo mass
(NFW;  Eke \etal  2001;  Jing  2002; Bullock  \etal  2001; Zhao  \etal
2003a,b).  We use the following $c$-$M$ relation:
\begin{equation}\label{concent}
c(M)=A \left(\frac{M}{M^*}\right)^{-0.13}\,,
\end{equation}
where    $M^*$   is    the   nonlinear    mass   scale    defined   as
$\sigma(M^*)=\delta_{\rm c}$. In most of our analyses we assume $A=14$, 
but we also test the sensitivity of our results to changes in $A$.

\subsection{Halo clustering}
\label{sec:haloclustering}

The number  density and density  profiles of dark matter  halos allow
one  to compute  the clustering  properties  of dark  matter on  small
scales. On large scales,  however, one needs information regarding the
spatial distribution  of dark  matter halos. The  halo-halo two-point
correlation  function  is related  to  the  mass correlation  function
through the  so-called halo bias factor  $b$ (e.g., Mo  \& White 1996,
2002; Sheth, Mo \& Tormen 2001). On large scale, we can write
\begin{equation}
\label{xihalo} 
\xi_{\rm hh}(r;M_1,M_2) = b(M_1) \, b(M_2) \, \xi_{\rm 2h}^{\rm dm}(r)\,,
\end{equation}
where $\xi_{\rm 2h}^{\rm dm}(r)$ is the 2-halo term of the dark matter
correlation, to be specified later, and
\begin{eqnarray}
\label{bm} 
b(M) & = & 1 + {1\over\sqrt{a}\delta_{\rm c}} \
\Bigl[ \sqrt{a}\,(a\nu^2) + \sqrt{a}\,b\,(a\nu^2)^{1-c} - \nonumber \\
& & {(a\nu^2)^c\over (a\nu^2)^c + b\,(1-c)(1-c/2)}\Bigr],
\end{eqnarray}
with $a=0.707$, $b=0.5$, $c=0.6$ and $\nu = \delta_{\rm c}/ \sigma(M)$
(Sheth, Mo \& Tormen 2001).  

We follow a similar  approach for the three-point correlation function
of dark matter halos and assume that it has the form
\begin{eqnarray}\nonumber
\zeta_{\rm hhh}(r_{12},r_{23},r_{31},M_1,M_2,M_3) &=&
b(M_1)b(M_2)b(M_3) \\
&\times &  \zeta_{\rm 3h}^{\rm dm}(r_{12},r_{23},r_{31})\,,
\end{eqnarray}
where  $\zeta_{\rm 3h}^{\rm  dm}(r_{12},r_{23},r_{31})$ is  the 3-halo
term of the 3PCF of dark matter (to be specified below).  Note that we
neglected  the quadratic  term  in the  relation  between halo  number
density and  mass density,  i.e. we assumed  linear bias  between halo
distribution and mass distribution, which  is expected to be valid in
quasi-linear  regime (e.g. Mo,  Jing \&  White 1997;  Ma \&  Fry 2000;
Takada \& Jain 2003). We will discuss the impact of including
this quadratic term on the 3PCF later in Section~\ref{sec:quad}.

\subsection{Halo occupation numbers}

In order to construct a model for the three-point correlation function
of galaxies, we need to know how galaxies populate dark matter halos.
Here  the  key  quantity  is  the  halo  occupation  number,  $\langle
N(M)\rangle$,  which describes  the average  number of  galaxies (with
luminosities greater than some limiting luminosity) that occupy a halo
of mass $M$. As discussed in Section~\ref{sec:intro}, numerous studies
have  used  such halo  occupation  number  models  to investigate  how
changes in $\langle N(M) \rangle$ impact on the statistical properties
of the galaxy distribution. In a  series of recent papers, Yang, Mo \&
van den Bosch  (2003) and van den Bosch, Yang \&  Mo (2003) have taken
this  halo occupation  approach one  step further  by  considering the
occupation  as  a  function  of  galaxy  luminosity  and  type.   They
introduced the conditional luminosity function (hereafter CLF) $\Phi(L
\vert  M)  {\rm  d}L$,  which   gives  the  number  of  galaxies  with
luminosities in the range $L \pm  {\rm d}L/2$ that reside in halos of
mass $M$.  Yang \etal  (2004) constructed mock galaxy redshift surveys
based on this CLF, and showed that many of the corresponding low-order
clustering  properties   are  in   good  agreement  with   the  2dFGRS
observations, both in real and redshift space.

With the  CLF the  occupation numbers can  be computed as  function of
both  luminosity and type.  For example,  the average  halo occupation
number for galaxies within a given luminosity range, $L_1<L<L_2$, is,
\begin{equation}
\label{meanN}
{\cal N}(M) = \int_{L_1}^{L_2} \Phi(L \vert M) \, {\rm d}L .
\end{equation}
This halo  occupation number  can be further  divided into  early- and
late-type components (e.g., van den Bosch \etal 2003),
\begin{equation}
{\cal N}(M) = {\cal N}_{\rm early}(M)+ {\cal N}_{\rm late}(M)\,.
\end{equation}
or into central and satellite  galaxy components (Yang \etal 2004; van
den Bosch \etal 2004),
\begin{equation}
{\cal N}(M) = {\cal N}_{\rm c}(M)+ {\cal N}_{\rm s}(M)\,,
\end{equation}
Here ${\cal  N}_{\rm c}$ is either zero or unity, and the central galaxy is
always located at the center of the halo. Satellite galaxies, on the
other hand, are assumed to follow a number density distribution given
by $n_{\rm s}(r)$. In this paper, we adopt the same CLF 
as the fiducial model in Yang \etal (2004), i.e. model D in 
van den Bosch \etal (2003).

\section{Halo-based models of two-point correlation functions}
\label{sec:2pcf}

In the halo model, the  two-point correlation function for dark matter
(and galaxies) can be decomposed into two parts,
\begin{equation}
\xi(r) = \xi_{\rm 1h}(r) + \xi_{\rm 2h}(r)\,,
\label{2pcf}
\end{equation}
where $\xi_{1  {\rm h}}$  represents the correlation  due to  pairs of
dark matter particles (or pairs of galaxies) within the same halo (the
``1-halo'' term), and $\xi_{2  {\rm h}}$ describes the correlation due
to dark matter particles  (galaxies) that occupy different halos (the
``2-halo'' term).   

For convenience, we introduce the normalized halo profile $u_{\rm M}(r)=\rho
(r)/M$, and  the normalized  number density distribution  of satellite
galaxies $u_{\rm s}(r)=n_{\rm s}(r)/{\cal N}_{\rm s}(M)$, so that
\begin{equation}
\int_{V_{\rm 180}} d^3{\bf x}~ u_{M,s}(r)=1\,,
\end{equation}
where  $V_{\rm 180}$ is the  volume of  the sphere  defined by  the virial
radius $r_{\rm 180}$. 

\subsection{Two-point correlation function for dark matter}
\label{sec:1h-2pcfdm}

The 1-halo term  mass correlation function can be  calculated from the
dark matter density distribution (Ma \& Fry 2000):
\begin{eqnarray}
\label{xidm1h}
\xi_{\rm 1h}^{\rm dm}(r) &=&  \frac{1}{{\overline \rho} ^2} \int_0^{\infty}
{\rm d}M \, n(M) \, \int_{V_{\rm 180}} d^3{\bf x}~ \rho(x)\rho(|{\bf
x + r}|) \nonumber \\
&=& \frac{1}{{\overline \rho} ^2} \int_0^{\infty} 
{\rm d}M \, n(M) \, M^2 \, f_{M}(r)
\end{eqnarray}
where  $f_{\rm M}(r)$ is the  particle pair  distribution function  within a
dark matter halo of mass $M$:
\begin{equation}
\label{eq:fs}
f_{\rm M}(r) = 2\pi \int_0^{r_{\rm 180}}u_{\rm M}(s)\, s^2ds \int^\pi_0
u_{\rm M}(|{\bf s}+{\bf r}|) \sin \theta\, d\theta\,,
\end{equation}
with  $|{\bf s}+{\bf r}|=(s^2+r^2+2sr\cos{\theta})^{1/2}$.

Formally, one can write the 2-halo term of the dark matter 2PCF as
\begin{eqnarray}
\label{eq:xi2hdm}
\xi_{\rm 2h}^{\rm dm}(r) &=& {1\over \bar{\rho}^2}
\int_0^{\infty} {\rmd M}_1 \, n(M_1) \, M_1
\int_0^{\infty} {\rmd M}_2 \, n(M_2) \, M_2 \nonumber \\
 & & \xi_{\rm hh}(r; M_1, M_2)\,. 
\end{eqnarray}
The  halo-halo correlation function  can be  directly related  to the
non-linear  2PCF $\xi_{\rm  NL}^{\rm  dm}(r)$ of  the  dark matter  by
taking account of halo-halo exclusion  and of the fact that dark matter
halos are biased tracers of the mass:
\begin{equation}
\xi_{\rm hh}(r; M_1, M_2) = b(M_1) b(M_2) U(r, M_1) U(r, M_2) 
\xi_{\rm NL}^{\rm dm}(r)\,,
\end{equation}
where 
\begin{equation}
U(r,M) = \left \{ \begin{array}{l} 0 \quad {\rm if} \quad r < r_{\rm exc}(M) \\
1 \quad {\rm else} \,.\end{array} \right .\,.
\end{equation}
Thus, we can write
\begin{equation}
\label{exclusion}
\xi_{\rm 2h}^{\rm dm}(r) = \left[f_{\rm exc}^{\rm dm}(r)\right]^2 \, 
\xi_{\rm NL}^{\rm dm}(r)\,,
\end{equation}
where
\begin{equation}
f_{\rm exc}^{\rm dm}(r)= {1 \over \overline{\rho}} \int_{0}^{\infty}
n(M) \, M \, b(M)\, U(r,M)\, {\rm d}M\,.
\end{equation}
We compute  $\xi^{\rm dm}_{\rm NL}(r)$  from the Fourier  transform of
the  non-linear power spectrum  $P_{\rm NL}(k)$  given by  Smith \etal
(2003).  As  we  show below,  the   choice  of   $r_{\rm exc}(M)$  can
significantly affect the amplitude  of the 2PCF on intermediate scales
($r\sim 2\mpch$).   Unfortunately, this  effect is difficult  to model
from first  principle.  What we will do  is to tune its  value so that
the model  prediction for  the 2PCF best  matches simulation  results. 
Note that when we consider the 2-halo term of the 2PCF, we do not take
into  account  the  impact  of  the  halo  profile.   This  effect  is
negligible compared to the halo-halo exclusion effect.

\begin{figure*}
\centerline{\psfig{figure=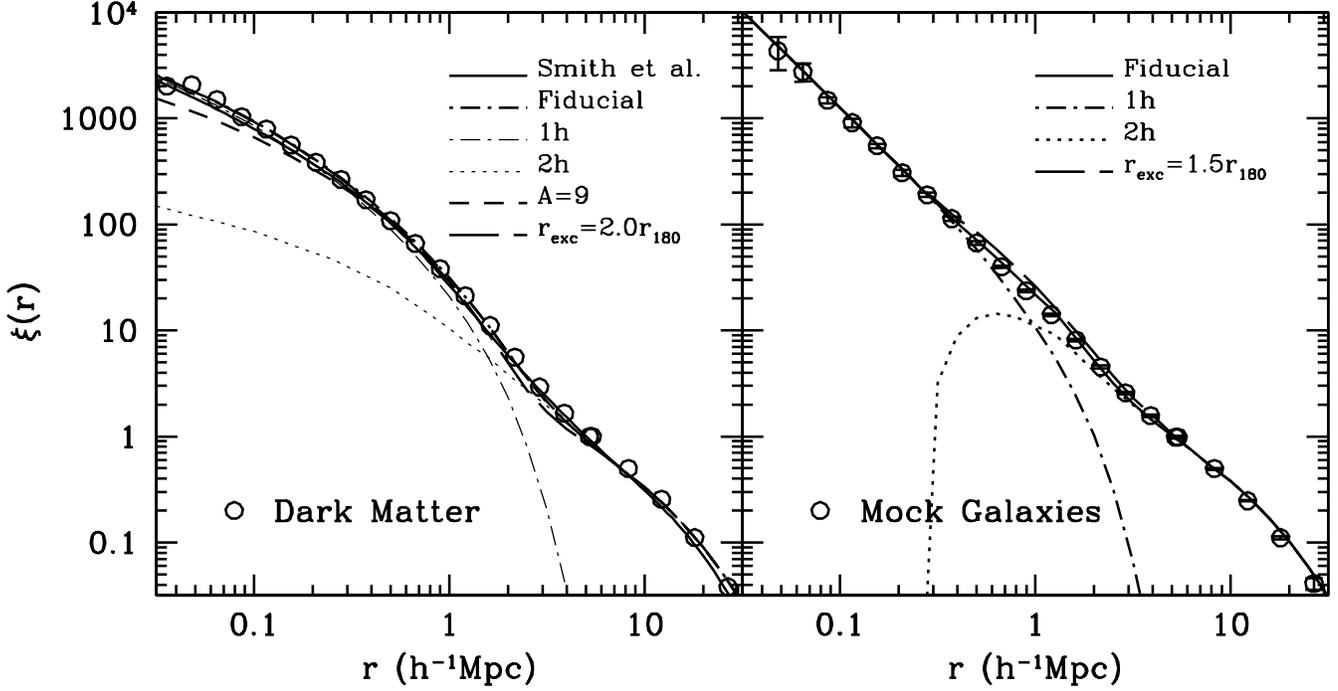,width=\hdsize}} 
\caption{The 2PCFs for dark matter (left panel) and mock galaxies with 
  luminosity $M_{b_J}- 5 \log h  < -18.5 $ (right panel).  The circles
  in the left  panel are the results obtained  from one $300\mpch$ box
  simulation; the  circles with error-bars  in the right panel  are the
  mean and 1-$\sigma$ error  obtained from four simulations.  The thin
  dot-dashed, thin doted and thick dot-dashed curves in the left panel
  are the  1-halo term, 2-halo  term and total halo  model predictions
  for  the fiducial  dark matter  2PCFs.  In  this fiducial  model, we
  adopted a  concentration normalization  $A=14$ and a  mean halo-halo
  exclusion radius  $r_{\rm exc} = 1.5~r_{\rm  180}$.  The short  and long
  dashed lines in the left panel correspond to the 2PCFs for $A=9$ and
  $r_{\rm exc} =  2.0~r_{\rm 180}$ models,  respectively.  For comparison,
  the fitting result of Smith \etal (2003) is shown as the solid line.
  The fiducial model predicts a  2PCF that is consistent with both the
  simulation and Smith  \etal results. In the right  panel, we compare
  the model prediction and simulation results for galaxies.  While for
  the galaxy 2PCFs,  the model with a mean  halo-halo exclusion radius
  $r_{\rm exc}  = 2.0~r_{\rm  180}$ consists  with the  simulation results
  extremely well (see the text for details).}
\label{fig:2pcf}
\end{figure*}

\subsection{Two-point correlation function for galaxies}
\label{sec:1h-2pcfg}

As for the dark matter, we split the 2PCF for galaxies in a 1-halo and
a 2-halo term.  In order to  model the 1-halo term, we need to specify
the distribution of galaxies in individual halos. As stated above, we
assume that the central galaxy is located at the halo center, and that
the satellite galaxies follow a normalized number density distribution
given  by  $u_{\rm s}(r)$.    For  simplicity,  unless  specifically  
stated otherwise, we  assume that  $u_{\rm s}(r) = u_{\rm M}(r)$, i.e., 
that  the number
density of satellite  galaxies is the same as that  of the dark matter
particles within the halos.

We can write the 1-halo term of the galaxy correlation function as
\begin{equation}\label{xi1h}
\xi_{\rm 1h}^{\rm g}(r) = \frac{2}{{\overline n}_{\rm g}^2}
\int_0^{\infty} n(M) \, {\langle N_{\rm pair}(M)\rangle} \,
f(r) \, {\rm d} M \,,
\end{equation}
where $\langle N_{\rm pair}(M) \rangle$ is the mean number of pairs in
halos of mass  $M$, $f(r) 4\pi r^2 \Delta r$ is  the fraction of pairs
with separation in the range $r\pm \Delta r/2$, and $\overline{n}_{\rm
  g}$ is the mean number density of galaxies given by
\begin{equation}
\label{barng} {\overline n}_{\rm g} = \int_{0}^{\infty} n(M) \,
{\cal N}(M) \, {\rm d}M \,.
\end{equation}
The mean  number of pairs  as function of separation,  $\langle N_{\rm
  pair}\rangle   f(r)$,  can  be   divided  into   contributions  from
central-satellite pairs and satellite-satellite pairs:
\begin{equation}
\label{eq:Npairfr}
\langle N_{\rm pair}\rangle f(r) =
\langle N_{\rm cs}\rangle u_{\rm s}(r) +
\langle N_{\rm ss}\rangle f_{\rm s}(r)\,,
\end{equation}
where $f_{\rm s}(r)$ follows from eq.~(\ref{eq:fs}) upon  substituting
$u_{\rm s}$ for $u_{\rm M}$. The number of central-satellite pairs is
\begin{equation}
\langle N_{\rm cs} \rangle = {\cal N}_{\rm c}(M) {\cal N}_{\rm s}(M)\,.
\end{equation}
Since  $\langle  N_{\rm  ss}\rangle$  depends  not only  on  the  mean
occupation ${\cal N}_{\rm s}(M)$,   but also  on the second moment, we 
adopt the nearest integer model in  which  $N_{\rm s}(M)$ has  the  
probability of  $N+1-{\cal  N}_{\rm s}(M)$  to  take the  value  $N$  
and  the probability  of  ${\cal  N}_{\rm s}(M)-N$ to take  the value 
$N+1$, if $N<{\cal  N}_{\rm s}(M) <N+1$.  In
this case, the mean number of satellite-satellite pairs is
\begin{equation}
\langle N_{\rm ss} \rangle = N \, {\cal N}_{\rm s}(M) - {1 \over 2} \,
 N \, (N+1) \,.
\end{equation}
The   2-halo   term   of   the   2PCF  for   galaxies   follows   from
eq.~(\ref{eq:xi2hdm})  upon  substituting  $\overline{n}_{\rm g}$  for
$\overline{\rho}$ and  ${\cal N}(M_1)$  and ${\cal N}(M_2)$  for $M_1$
and $M_2$, respectively. This yields
\begin{equation}
\xi_{\rm 2h}^{\rm g}(r) = \left[f_{\rm exc}^{\rm g}(r)\right]^2 \, 
\xi_{\rm NL}^{\rm dm}(r)\,,
\end{equation}
where
\begin{equation}
f_{\rm exc}^{\rm g}(r)= {1 \over \overline{n}_{\rm g}}
\int_{0}^{\infty} n(M) \, {\cal N}(M) \, b(M)\, U(r,M)\, {\rm d}M\,.
\end{equation}
As for the  dark matter, we consider $r_{\rm exc}$  a   free parameter 
which we tune to best match the 2PCF of our mock galaxies.

\section{Halo-based models of three-point correlation functions}
\label{sec:3pcf}

The three-point  correlation function $\zeta(r_{12},  r_{23}, r_{31})$
in real space  is defined through the probability  ${\rm d}P_{123}$ of
finding  one  particle simultaneously  in  each  of  the three  volume
elements ${\rm d}V_1$, ${\rm d}V_2$  and ${\rm d}V_3$ that are located
at ${\bf  r}_{1}$, ${\bf r}_{2}$ and ${\bf  r}_{3}$, respectively.  By
definition, this probability is related to the 3PCF as
\begin{eqnarray}
\label{eq:3pcf1} {\rm d}P_{123} &=& \bigl[ 1 +\xi(r_{12})
+\xi(r_{23}) + \xi(r_{31}) +
\zeta(r_{12},r_{23},r_{31}) \bigr] \nonumber \\
 & & \times  {\bar n}^3 {\rm d}V_{1} \, {\rm d}V_{2} \, {\rm d}V_{3}\,,
\end{eqnarray}
where $r_{ij}=|{\bf  r}_{i}-{\bf r}_{j}|$  and $\bar n  $ is  the mean
number density of particles (Peebles  1980).  It is common practice to
express the 3PCF in the so-called normalized form,
\begin{equation}
Q(r,u,v) = {\zeta(r_{12},r_{23},r_{31}) \over \xi(r_{12})\xi(r_{23})
+\xi(r_{23})\xi(r_{31}) +\xi(r_{31})\xi(r_{12})} \,,
\label{eq:Qruv}
\end{equation}
where, following Peebles (1980), the new variables,
\begin{equation}
\label{ruv} 
r =   r_{12}, \hskip1cm 
u = {{r_{23}}\over{r_{\rm 12}}}, \hskip1cm 
v = {{r_{31}-r_{23}}\over{r_{12}}}\,,
\end{equation}
describe the  shape ($u$ and $v$)  and size ($r$) of  the triplet with
sides $r_{12}<r_{23}<r_{31}$.  If $Q(r,u,v)$  is constant, the 3PCF is
said to have the `hierarchical form', i.e.
\begin{equation}
\label{zetarrr} 
\zeta(r_{1},r_{2},r_{3}) \propto 
[\xi(r_{1})\xi(r_{2}) + \xi(r_{2})\xi(r_{3}) + \xi(r_3)\xi(r_1)]\,.
\end{equation}

Following the approach  for the 2PCF, we write the 3PCF  as the sum of
the `1-halo', `2-halo' and `3-halo' terms:
\begin{eqnarray}\nonumber
\zeta(r_{12},r_{23},r_{31})&=&\zeta_{\rm 1h}(r_{12},r_{23},r_{31})
        +\zeta_{\rm 2h}(r_{12},r_{23},r_{31})  \\
       &+& \zeta_{\rm 3h}(r_{12},r_{23},r_{31})\,, 
\label{3pcf}
\end{eqnarray}
where $r_{12}$, $r_{23}$,  and $r_{31}$ are the lengths  for the three
edges of the triplet. Without loosing generality, we assume $r_{12}\le
r_{23}\le r_{31}$. In the remainder of this section, we present models
for the various terms for both dark matter particles and galaxies.

\begin{figure}
\centerline{\psfig{figure=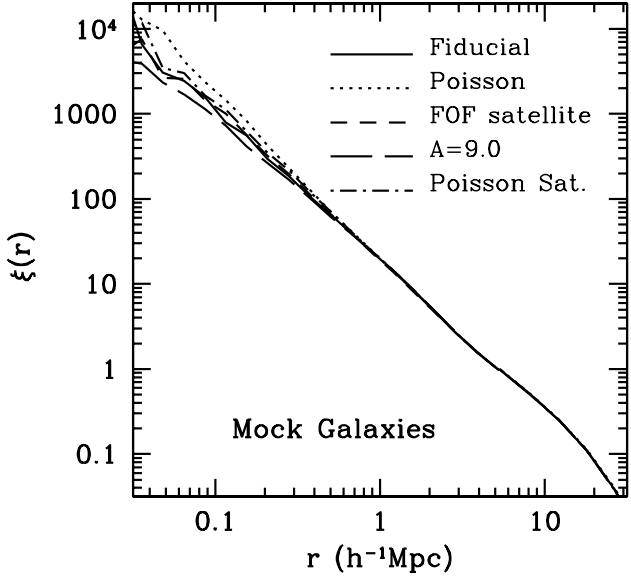,width=\hssize}}
\caption{The 2PCFs for mock galaxies with different models for
  the  satellite  distribution  in  halos.   The solid  line  is  the
  fiducial model,  where satellite distribution in a  halo follows the
  NFW with  the concentration  normalization $A=14$.  The  dotted line
  corresponds to result  where the number of galaxies in  a halo has a
  Poisson distribution with the mean given by the mean halo occupation
  number.   The  short  dashed  line  (FOF) is  the  result  in  which
  satellite galaxies  are traced by  dark matter particles in  the FOF
  halos that host  them.  The long dashed line  shows the result with
  the  concentration  normalization   $A=9$. The dot-dashed line
  is the result where the number of {\it satellite galaxies} in a halo 
  follows a Poisson distribution with the mean given by the mean 
  occupation number of satellite galaxies.
  Due  to  the  simulation resolution,  only galaxies  with 
  $M_{b_J}-  5  \log h  < -18.5$  are used.}
\label{fig:2pg-fof}
\end{figure}

\subsection{The three-point correlation function for
dark matter particles}

Following the method described in  Takada \& Jain (2003), we write the
1-halo term of the 3PCF for dark matter as
\begin{equation}
\zeta_{\rm 1h}^{\rm dm}(r_{12},r_{23},r_{31})=\frac{1}{{\overline
\rho}^3} \int {\rm d}M \, n(M) \, M^3 \, g_{M}(1,2,3)\,,
\end{equation}
where
\begin{eqnarray}\nonumber
g_{\rm M}(1,2,3) & = & \int_0^{r_{\rm 180}} u_{\rm M}(s)\, s^2 d s
\int_0^{\pi} u_{\rm M}(|{\bf s} + {\bf r_{12}}|) \sin{\theta}\, d \theta
\\
&\times&  \int_0^{2\pi} u_{\rm M}(|{\bf s} + {\bf r_{13}}|)\, d
\varphi\,. \label{eq:gs}
\end{eqnarray}
Here $|{\bf s} + {\bf r}_{12}| = (s^2 + r_{12}^2 +
2sr_{12} \cos{\theta} )^{1/2}$  and $|{\bf s} + {\bf
r}_{13}|=(s^2+r_{13}^2 +2sr_{13}\cos{\theta_1})^{1/2}$.
The angle $\theta_1$ is given by 
\begin{equation}
\cos{\theta_1}=\sin{\alpha}\sin{\theta}\cos{\varphi}
+\cos{\alpha}\cos{\theta}\,,
\end{equation}
with 
\begin{equation}
\alpha = \cos^{-1} \left[
\frac{r_{12}^2+r_{31}^2-r_{23}^2}{2r_{12}r_{31}} \right]\,.
\end{equation}

The 2-halo term in the 3PCF of dark matter can be written as
\begin{eqnarray}
\zeta_{\rm 2h}^{\rm dm}(r_{12},r_{23},r_{31}) &= & \epsilon_{\rm
1h}^{\rm dm} (r_{12}) \frac { \xi_{\rm 2h}^{\rm dm} (r_{23}) +
\xi_{\rm 2h}^{\rm dm} (r_{31})}{2} \nonumber \\
&+& {\rm perm}(1,2,3)\,,
\end{eqnarray}
where ${\rm perm}(1,2,3)$ is the permutation of the three points, and
\begin{equation}
\label{3pcf2h}
\epsilon_{\rm 1h}^{\rm dm} (r) = \frac{1}{{\overline \rho}^2 }
\int_0^{\infty} {\rm d}M \, n(M) \, M^2 \, b(M) \, f_{\rm M}(r) \,,
\end{equation}
with $f_{\rm M}(r)$ defined in eq.~(\ref{eq:fs}).

Similar  as with the  2-halo term  in the  2PCF, we  can use  the halo
exclusion principle to  write the 3-halo term of the  3PCF in terms of
the  non-linear, dark  matter  3PCF, $\zeta_{\rm  NL}^{\rm  dm}$ (cf.  
eq.~[\ref{exclusion}]).   However, no  accurate theoretical  model for
$\zeta_{\rm NL}^{\rm dm}$  that can cover both small  and large scales
is  currently  available.  We  therefore use  the  quasi-linear  3PCF,
$\zeta_{\rm NL}^{\rm dm}$, instead, and write
\begin{eqnarray}
\label{zeta3hdm}
\zeta_{\rm 3h}^{\rm dm}(r_{12},r_{23},r_{31}) &=& 
\left[f_{\rm exc}^{\rm dm}(r_{12}) \, f_{\rm exc}^{\rm dm}(r_{23}) \, 
f_{\rm exc}^{\rm dm}(r_{31})\right] \, \nonumber \\
&\times& \zeta_{\rm QL}^{\rm dm} (r_{12},r_{23},r_{31})\,.
\end{eqnarray}
Note that  the quasi-linear 3PCF is  only applicable at  large scales. 
However,  since the  number of  triplets on  small,  highly non-linear
scales is expected to be dominated by the 1-halo and 2-halo terms, the
use of for $\zeta_{\rm NL}^{\rm dm}$ to compute the 3-halo term should
be sufficiently accurate. The quasi-linear 3PCF has been obtained from
perturbation theory (e.g.   Fry 1984; Matsubara \& Suto  1994; Jing \&
B\"orner 1997; Barriga \& Gazta\~{n}zga 2002), and can be written as
\begin{eqnarray}
\lefteqn{\zeta_{\rm QL}^{\rm dm}(r_{12},r_{23},r_{31}) =
\frac{10}{7}\xi(r_{12})\xi(r_{23}) + \nonumber }\\
\lefteqn{+\frac{4}{7}
\left\{-3\frac{\phi'(r_{12})\phi'(r_{23})}{r_{12}r_{23}}
-\frac{\xi(r_{12})\phi'(r_{23})}{r_{23}}
-\frac{\xi(r_{23})\phi'(r_{12})}{r_{12}}\right. \nonumber }\\
\lefteqn{\left. +\mu ^2\left[\xi(r_{12})+
3\frac{\phi'(r_{12})}{r_{12}}\right]\left[\xi(r_{23})
+3\frac{\phi'(r_{23})}{r_{23}}\right] \right\} \nonumber} \\
\lefteqn{-\mu\left[\xi'(r_{12})\phi'(r_{23})+\xi'(r_{23})\phi'(r_{12})\right]
+ {\rm perm}\, (1,2,3) \,,}
\label{eq:zetaql}
\end{eqnarray}
where $\mu = ({\bf r}_{12}\cdot {\bf r}_{32})/(r_{12}r_{23})$,
\begin{equation}
\phi(r)=\frac{1}{2\pi^2}\int_0^{\infty} \frac{P_{\rm L}
(k)}{k^2}\frac{\sin(kr)}{kr} \, k^2 {\rm d}k \,,
\end{equation}
with  $P_{\rm  L}  (k)$  the  linear  power  spectrum,  $\phi'(r)={\rm
  d}\phi/{\rm d}r$, and ${\rm  perm}\, (1,2,3)$ the permutation of the
three  points of the  triplets.  $N$-body  simulations show  that this
formula is a good approximation on scales $r_{12}\ga 6 \mpch$ (Jing \&
B\"orner 1997; Barriga \&  Gazta\~naga 2002).  Since we are interested
in the  reduced 3PCF $Q(r_{12},r_{23},r_{31})$, which we  define to be
the  3PCF normalized  by the  square of  the nonlinear  2PCF $\xi_{\rm
  NL}^{\rm   dm}(r)$,    we   made   a    modification   in   equation
(\ref{eq:zetaql})  by  replacing  $\xi  (r)$ with  $\xi_{\rm  NL}^{\rm
  dm}(r)$.

\subsection{The three-point correlation function for galaxies}

As  for the  dark matter,  we write  the 1-halo  term in  the  3PCF of
galaxies as
\begin{equation}
\zeta_{\rm 1h}^{\rm g}(r_{12},r_{23},r_{31})=\frac{6}{{\overline
n}_{\rm g}^3} \int_{0}^{\infty} {\rm d}M \, n(M) \,
\langle N_{\rm triplet} \rangle \, g(1,2,3)\,,
\end{equation}
where  $\langle  N_{\rm  triplet}\rangle$  is the  average  number  of
triplets  per halo  of  mass $M$  and  $g(1,2,3)$ is  the fraction  of
triplets in a particular configuration. Note that both $\langle N_{\rm
  triplet}\rangle$  and $g(1,2,3)$ may  depend on  halo mass  $M$. The
triplets   can  be   divided   into  central-satellite-satellite   and
3-satellite triplets:
\begin{equation}
\langle N_{\rm triplet}\rangle g(1,2,3) =  \langle N_{\rm
css}\rangle g_{\rm c}(1,2,3) + \langle N_{\rm sss}\rangle g_{\rm s}(1,2,3)\,,
\end{equation}
where $g_{\rm s}(1,2,3)$ follows  from eq.~(\ref{eq:gs}) upon substituting
$u_{\rm M}$ with $u_{\rm s}$, and $g_{\rm c}(1,2,3)$ is given by
\begin{equation}
g_{\rm c}(1,2,3) = 
\frac{1}{3}\left[u_{\rm s}(r_{12})\, u_{\rm s}( r_{23}) +{\rm
perm} (1,2,3)\right]\,.
\end{equation} 
Using the same sampling algorithm as described in Section
~\ref{sec:1h-2pcfg} ( $N_{\rm s}(M)$ has  the  probability of  
$N+1-{\cal  N}_{\rm s}(M)$  to  take the  value  $N$  and  the probability  
of  ${\cal  N}_{\rm s}(M)-N$ to take  the value $N+1$, if 
$N<{\cal  N}_{\rm s}(M) <N+1$),   we   can    write    the   number    of
central-satellite-satellite triplets, $\langle N_{\rm css}\rangle$, as
\begin{equation}
\langle N_{\rm css} \rangle = {\cal N}_{\rm c}(M)\, \langle N_{\rm ss} 
\rangle \,,
\end{equation}
and the number of  3-satellite triplets, $\langle N_{\rm sss}\rangle$, as
\begin{equation}
\langle N_{\rm sss} \rangle = { N(N-1){\cal N}_{\rm s}(M) \over 2}
- {N(N^2-1) \over 3}\,.
\end{equation}

The 2-halo  term in the 3PCF for  galaxies is similar to  that of dark
matter particles, and can be expressed as
\begin{eqnarray}\label{3pcf2h}
\zeta_{\rm 2h}^{\rm g}(r_{12},r_{23},r_{31}) &=&  \epsilon_{\rm
1h}^{\rm g}(r_{12}) \frac {\xi_{\rm 2h}^{\rm g}(r_{23}) +
\xi_{\rm 2h}^{\rm g} (r_{31})}{2} \\
 &+& {\rm perm}(1,2,3)\,, \nonumber
\end{eqnarray}
where
\begin{equation}
\epsilon_{\rm 1h}^{\rm g} (r) = \frac{1}{{\overline n}_{\rm g}^2
\, {\overline b}} \int_0^{\infty} {\rm d}M \, n(M) \, b(M)\,
\langle N_{\rm pair}\rangle \, f(r)\,,
\end{equation}
with   $f(r)$  given by eq.~(\ref{eq:Npairfr}) and
\begin{equation}
\label{averbias}
\overline{b} = {1 \over \overline {n}_{\rm g}}
\int_{0}^{\infty} n(M) \, {\cal N}(M) \, b(M) \, {\rm d} M\,.
\end{equation}

Finally, for the 3-halo term of the galaxy 3PCF we write
\begin{eqnarray}
\zeta_{\rm 3h}^{\rm g}(r_{12},r_{23},r_{31}) &=& 
\left[f_{\rm exc}^{\rm g}(r_{12})\, f_{\rm exc}^{\rm g}(r_{23}) \, 
f_{\rm exc}^{\rm g}(r_{31})\right] \, \nonumber \\
 & \times & \zeta_{\rm QL}^{\rm dm}(r_{12},r_{23},r_{31}) \,.
\end{eqnarray}

\begin{figure*}
\centerline{\psfig{figure=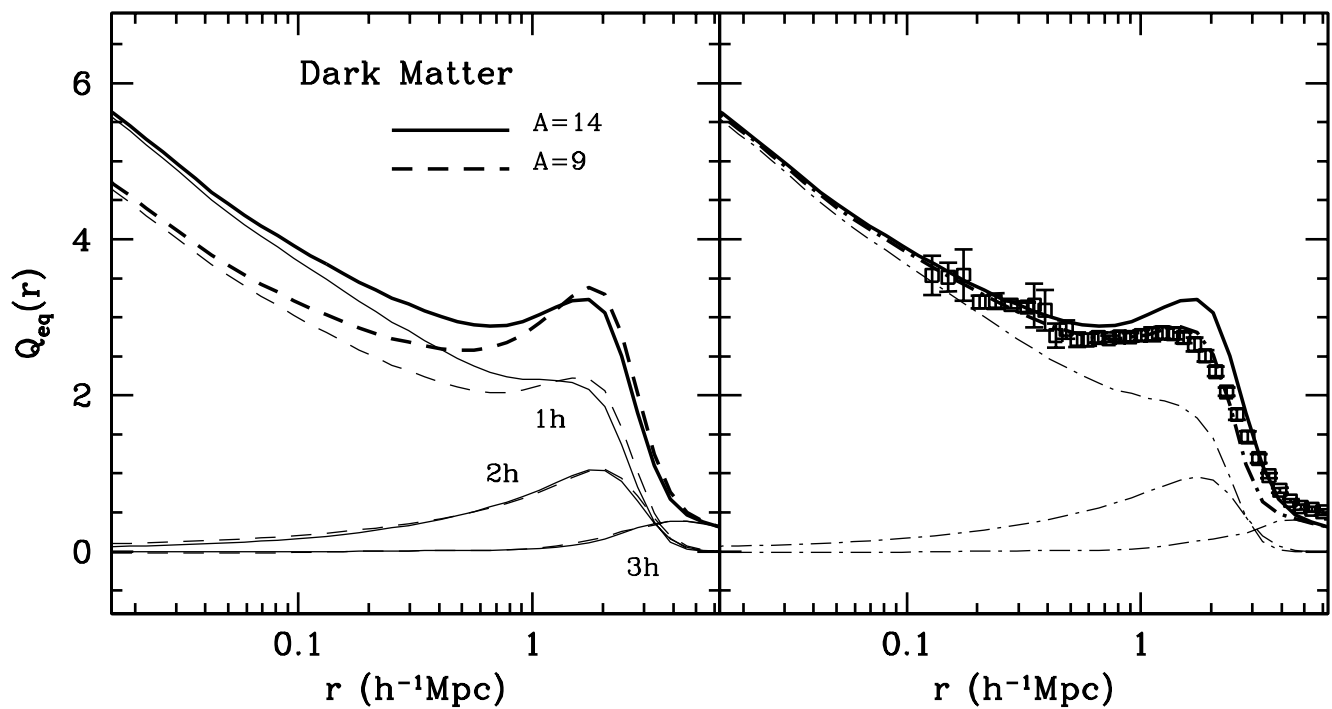,width=\hdsize}}        
\caption{The normalized  3PCFs of  equilateral  triangles for  dark  
  matter as  a function of  the size of  the triangle.  The  thick and
  thin curves  show the total and  1-, 2- and  3-halo contributions to
  the  3PCF, respectively.   The solid  and dashed  lines in  the left
  panel  are  results for  dark  matter  with different  concentration
  normalization  $A$.  We  compare model  predictions  with simulation
  results in the right panel.   Squares with error bars are simulation
  results.  The  thick solid  curve is  the same as  that in  the left
  panel, while  the dot-dashed lines  are the results that  assume the
  mass   function  in   the  model   is  truncated   at   $M=  2\times
  10^{15}\msunh$ to mimic the incompleteness in the simulations.}
\label{fig:3pdm-eq}
\end{figure*}

\section{Comparison with Simulations}
\label{sec:results}

The analytical  halo models for  the two- and  three-point correlation
functions of dark matter and galaxies are derived based on a number of
simple assumptions, and so their validity needs to be checked. In this
section, we use high-resolution  numerical simulations and mock galaxy
distributions (hereafter MGDs), to test the accuracy of these models.

The set  of $N$-body simulations  used here were  carried out by  Y.P. 
Jing and  Y. Suto (see  Jing 2002; Jing  \& Suto 2002) on  the VPP5000
Fujitsu  supercomputer  of the  National  Astronomical Observatory  of
Japan using a vectorized-parallel P$^3$M  code.  The set consists of a
total of  six simulations, each  of which used $N=512^3$  particles to
evolve the  distribution of  dark matter from  an initial  redshift of
$z=72$  down  to  $z=0$  in  a  $\Lambda$CDM  `concordance'  cosmology
($\Omega_m = 0.3$,  $\Omega_{\Lambda} = 0.7$, $H_0 =  70 \kmsmpc$, and
$\sigma_8  =  0.9$). All  simulations  consider  boxes with  periodic
boundary conditions. Two of them use boxsize
$L_{\rm box}=100 h^{-1} \Mpc$, with a force softening 
length of $\sim 10 h^{-1} \kpc$,  while the other  
four simulations have $L_{\rm  box}=300 h^{-1}  \Mpc$,  with 
force softening length $\sim 30 h^{-1} \kpc$. 
Different  simulations   with  the   same  box  size   are  completely
independent realizations and are used to estimate uncertainties due to
cosmic variance.  The particle masses are $6.2 \times 10^8 \msunh$ and
$1.7\times 10^{10}  \msunh$, for the small and large  box simulations,
respectively.

The mock  galaxy distributions (MGDs) are obtained  by populating dark
matter halos in these $N$-body simulations with galaxies according to
the conditional luminosity function  (CLF) model.  The construction of
such MGDs  is described in Yang  \etal (2004) and van  den Bosch \etal
(2004), and we refer the reader  to these papers for details..  We use
the same  CLF model  as in Yang  \etal (2004), which  yields excellent
fits  to the  observed LFs  and  the observed  correlation lengths  as
function of both  luminosity and type.  The satellite  galaxies in the
MGDs are  assumed to  be distributed within  $r_{\rm 180}$ and  with a
number density  distribution that follows the  density distribution of
the dark matter  (i.e.  $u_{\rm s}(r)=u_{\rm M}(r)$). In this section  
we only use
the $300 h^{-1}\Mpc$ simulations.   As discussed in Yang \etal (2004),
because  of the  finite resolution  of these  simulations,  the galaxy
population  in  these MGDs  are  only  complete  down to  an  absolute
magnitude limit of $M_{b_J}- 5 \log h \approx -18.5$.

\subsection{Two-point correlation functions}
\label{sec:2pcf-comp}

The  left-hand panel  of Figure~\ref{fig:2pcf}  plots the  dark matter
2PCFs obtained from the analytical halo model described above (various
lines) and  from the numerical simulations (open  circles).  The solid
line  corresponds to  the nonlinear  2PCF calculated  using  the Smith
\etal  (2003) fitting  formula,  and  is in  good  agreement with  the
simulations.  The  dot-dashed line  indicates the predictions  for our
fiducial model,  where we have assumed  $A=14$ and $r_{\rm exc}=1.5
~r_{\rm 180}$. With these assumptions, the predicted 2PCF matches both the
Smith  \etal model  and the  simulation results  extremely  well.  The
dashed  and  dotted curves  show  the  corresponding contributions  of
1-halo and 2-halo terms, respectively, and are shown for completeness.
We also plot the results for different halo concentrations ($A=9$) and
for a different exclusion radius  ($r_{\rm exc}=2.0~r_{\rm 180}$). The
former predicts  a significantly lower  1-halo term, while  the latter
predicts a lower 2-halo term on intermediate scales ($r\sim 2\mpch$).

The right panel of  Figure~\ref{fig:2pcf} shows the 2PCFs for galaxies
with  luminosity $M_{b_J}- 5  \log h<  -18.5$ where  we set  
$u_{\rm s}(r) = u_{\rm M}(r)$,  
i.e., the  number density  of galaxies  follows  the density
distribution  of the dark  matter.  The  lines are  model predictions,
while the open  circles with error-bars indicate the  mean and variance
obtained from four independent MGDs. This time the mean halo exclusion
radius  that  best matches  the  results  obtained  from the  MGDs  is
$r_{\rm exc} = 2.0~r_{\rm 180}$.   Although  in  good  agreement  with
Magliocchetti \& Porciani (2003),  this radius is slightly larger than
that used for the dark matter.  If we use the same exclusion radius as
for the dark matter, $r_{\rm exc} = 1.5~r_{\rm 180}$, the 2PCF of galaxies
is slightly  over-estimated on scales  $r\sim 2\mpch$.  The  fact that
the  model predictions are  sensitive to  the choice  of the  value of
$r_{\rm exc}$  suggests that  the predicting  power of  current halo-based
models are somewhat limited  around the transitional scale between the
1-halo and 2-halo components.

Using very  similar MGDs, Yang \etal  (2004) examined how  the 2PCF of
galaxies  depends on  the spatial  distribution of  satellite galaxies
within individual  dark matter halos.   Here for completeness  and to
compare with  such dependence  in the 3PCF  to be discussed  later, we
calculate the 2PCFs for different  $u_{\rm s}(r)$ and different sampling of 
galaxies in the dark matter halos.  The results are shown in
Fig~\ref{fig:2pg-fof}.  The solid line shows the 2PCF for the fiducial
model where $u_{\rm s}(r)  = u_{\rm M}(r)$ with $A=14$.  
For  comparison, the long
dashed  line  shows  the   results  with  $A=9$.   As  expected,  less
concentrated  distributions of  satellite galaxies  result  in reduced
correlation functions on small scales.  The dotted line corresponds to
a model  in which the  number of  galaxies in a  halo is drawn  from a
Poisson  distribution  with the  mean  given  by  the mean  occupation
number.   This results in  a significant  increase of  the correlation
power  on small  scales, and  is a  consequence of  the fact  that the
second moment of  the Poisson distribution is larger  than that of the
nearest integer  distribution adopted in our fiducial  model (see also
Benson  \etal  2000; Berlind  \&  Weinberg  2002;  Yang \etal  2003).  
In a recent study, Kravtsov \etal (2004) found that, while 
each halo may contain a central galaxy, the number of
satellite galaxies follows a Poisson distribution. For comparison,
we show the result of the MGD thus generated in Fig~\ref{fig:2pg-fof} 
as the dot-dashed line. This result of this model
agrees extremely well with that of the the fiducial model.
Finally, the short-dashed line indicates the 2PCF for a model in which
each satellite galaxy is assigned  the position of a randomly selected
particle from  the friends-of-friends (FOF) group  associated with the
halo under consideration. Note that  this yields a $\xi(r)$ that is in
excellent  agreement with  our fiducial  model.  Overall,  the spatial
distribution  of satellite  galaxies only  has a  small effect  on the
2PCF, and only on small scales ($r<0.5\mpch$).

\begin{figure}
\centerline{\psfig{figure=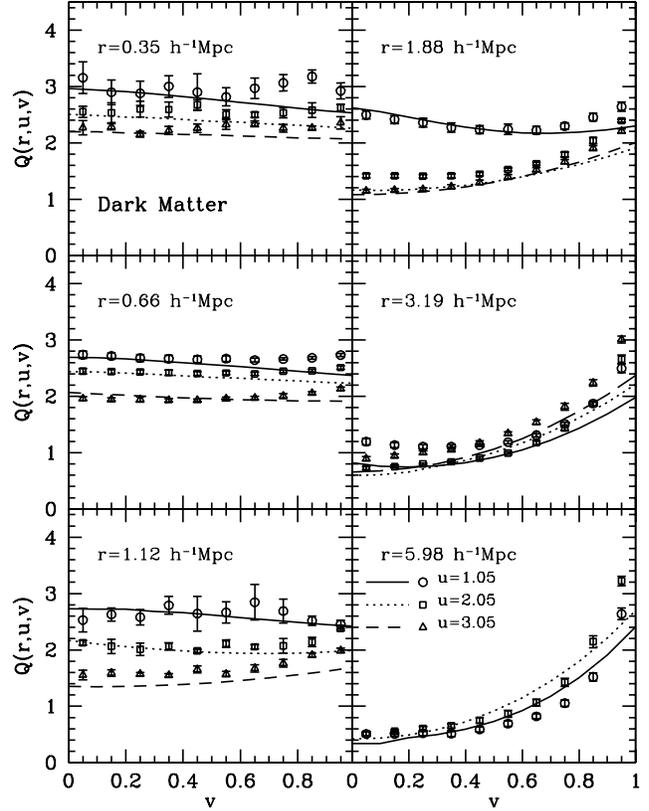,width=\hssize}} 
\caption{The normalized 3PCFs for dark matter. Symbols with 
  error-bars  are   simulation  results;  lines  are   the  halo  model
  predictions. }
\label{fig:3pdm}
\end{figure}

\subsection{The three-point correlation functions}
\label{sec:3PCF1}

To  compute  the  3PCFs  of  dark matter  particles  and  mock  galaxy
distribution we compare the counts  of triplets with those of randomly
distributed points:
\begin{eqnarray}
\label{eq:3pcf2} 
\zeta(r_{12},r_{23},r_{31}) &=& \frac {N_{\rm R}^3 \, 
DDD(r_{12},r_{23},r_{31})}
{N_{\rm D}^3 \, RRR(r_{12},r_{23},r_{31})} \nonumber \\
 & & - \xi(r_{12})- \xi(r_{23})- \xi(r_{31}) - 1\,,
\end{eqnarray}
Here $DDD$  and $RRR$ are the  triplet counts with  separations in the
ranges $r_{12}\pm  \Delta r_{12}/2$, $r_{23}\pm  \Delta r_{23}/2$, and
$r_{31}\pm  \Delta  r_{31}/2$, in  the  data  ($D$)  and random  ($R$)
samples,  respectively, and $N_{\rm D}$  and $N_{\rm R}$ correspond to  
the total number of objects in each sample.

We compute the normalized 3PCFs, $Q(r,u,v)$, for dark matter particles
and  mock  galaxies,  using  equal  logarithmic  bins  for  $r$,  with
$\Delta{\rm log} r \sim 0.05$, and  equal linear bins for $v$ and $u$,
with $\Delta  v = \Delta  u =0.1$.  For  the dark matter, we  use four
subsamples of about  500,000 particles, each selected from  one of the
four  realizations of  the $300  h^{-1}{\rm Mpc}$  box  simulations to
estimate $DDD$.  For  the MGDs, there are about  470,000 galaxies with
absolute magnitudes  $M_{b_J} - 5\log h  < -18.5$ in each  of the four
mock samples.   We use two sets  of random samples to  estimate $RRR$. 
The first contains 800,000 random  points in the simulation box and is
used to estimate the number of triplets with $5\mpch <r_{31}<20\mpch$.
The second  contains 10 times as  many points and is  used to estimate
the  counts of triplets  with $r_{31}\le  5\mpch$.  This  ensures that
$RRR  > 200$  for all  triangle configurations  of interest.   In what
follows, unless  specifically stated otherwise,  we adopt $r_{\rm exc}=1.5
\,r_{\rm  180}$  for dark  matter  particles and  $r_{\rm exc}=2.0\,r_{\rm
  180}$ for galaxies, and set $A=14$ and $u_{\rm s}(r) = u_{\rm M}(r)$.
\begin{figure}
\centerline{\psfig{figure=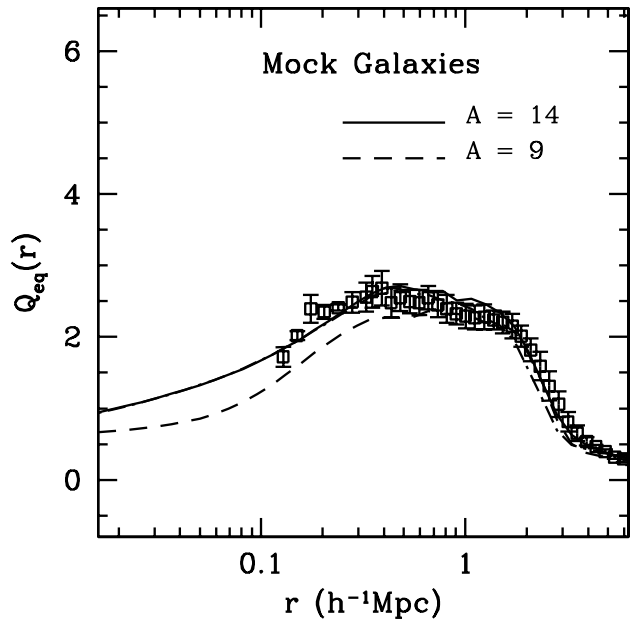,width=\hssize}}
\caption{The same as Figure~\ref{fig:3pdm-eq}, but here results
  are shown for  mock galaxies. Note that the  $Q_{eq}(r)$ of galaxies
  is extremely different  from that of dark matter  particles on small
  scales ($r<0.3 \mpch$). }
\label{fig:3pg-eq}
\end{figure}

\subsubsection{The 3PCF for dark matter}

We  first consider  the normalized,  dark matter  3PCF  of equilateral
($u=1$,  $v=0$) triangles,  $Q_{\rm eq}(r)$.   The left-hand  panel of
Figure~\ref{fig:3pdm-eq}  plots  $Q_{\rm  eq}(r)$ obtained  using  our
fiducial model with $A=14$ and $r_{\rm exc} = 1.5 r_{180}$. As shown in
Fig.~\ref{fig:2pcf}  this  normalization  of the  halo  concentrations
yields the best-fit to the 2PCF of the dark matter.  For comparison we
also show results  for a model with $A=9$.   As expected, reducing the
halo  concentration signicantly lowers  the value  of $Q_{\rm  eq}$ on
small scales.
  
The  right-hand  panel of  Fig.~\ref{fig:3pdm-eq}  compares the  model
predictions  with  the  simulation  results.  The  open  squares  with
errorbars are the  mean and 1-$\sigma$ variance of  the 3PCFs for dark
matter.   On intermediate  scales  of $r\sim  2\mpch$  the halo  model
predicts a  significantly higher $Q_{\rm  eq}$ than for  the numerical
simulations.  Since  the 1-halo term  of the 3PCF  $\zeta_{1h} \propto
M^3$,  one expects  $Q_{\rm eq}(r)$  to be  extremely senstive  to the
abundance of the most massive haloes. Due to the limited volume probed
by the $N$-body simulations, there  is a maximum halo mass above which
the  halo  mass function  is  no  longer  properly sampled.   For  the
simulations used  here, we estimate  this mass to be  $2\times 10^{15}
\msunh$. If we include an artifical  cut-off at this mass scale in our
theoretical mass function, we obtain  the 3PCF shown by the dot-dashed
curve  in the  right-hand  panel of  Fig.~\ref{fig:3pdm-eq}.  In  this
case, the  model prediction  matches the simulation  results extremely
well.   This  clearly demonstrates  that  the three-point  correlation
function  on  intermediate  scales   is  extremely  sensitive  to  the
abundance of massive haloes.

Fig.~\ref{fig:3pdm}  shows  a  comparison  of  model  predictions  and
simulation  results  for   non-equilateral  triangle  configurations.  
Symbols with errorbars correspond  to the $Q(r,u,v)$ obtained from the
numerical simulations, with lines  indicating the predictions from our
fiducial model, where we have included an upper limit to the halo mass
function  of $M  =2\times 10^{15}  \msunh$.  Without  this  limit, the
model  predictions  are  higher  by  about  20\%.   On  small  scales,
$Q(r,u,v)$ depends weakly  on $r$ and $v$, but  quite significantly on
$u$.    On  large   scales  ($r   >  3\mpch$),   $Q(r,u,v)$  increases
significantly  with increasing  $v$, which is consistent with previous
studies (i.e. Jing \& B\"orner 1997; Barriga \& Gazta\~naga 2002).
 Overall, the model  predictions agree  remarkable well  with the
simulation  results  on  almost   all  scales  and  for  all  triangle
configurations  considered,   indicating  that  the   halo  model,  as
presented here, is well suited to describe the 3PCF.

\subsubsection{The 3PCF for galaxies}
\label{sec:3pcfg}
\begin{figure}
\centerline{\psfig{figure=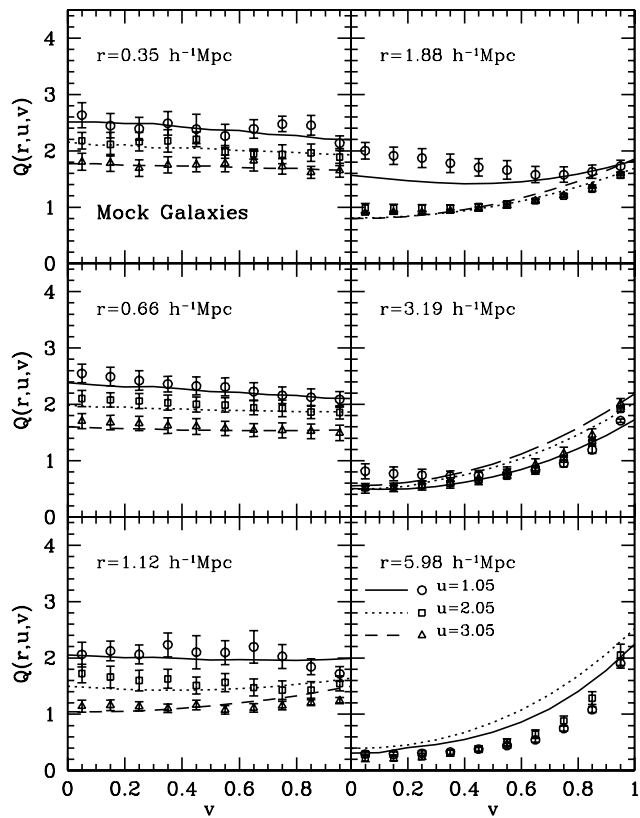,width=\hssize}}
\caption{The normalized 3PCFs for mock galaxies with
  $M_{b_j}- 5 \log  h < -18.5$. Symbols with  errorbars are simulation
  results; lines are the halo model predictions.}
\label{fig:3pg}
\end{figure}
\begin{figure*}
\centerline{\psfig{figure=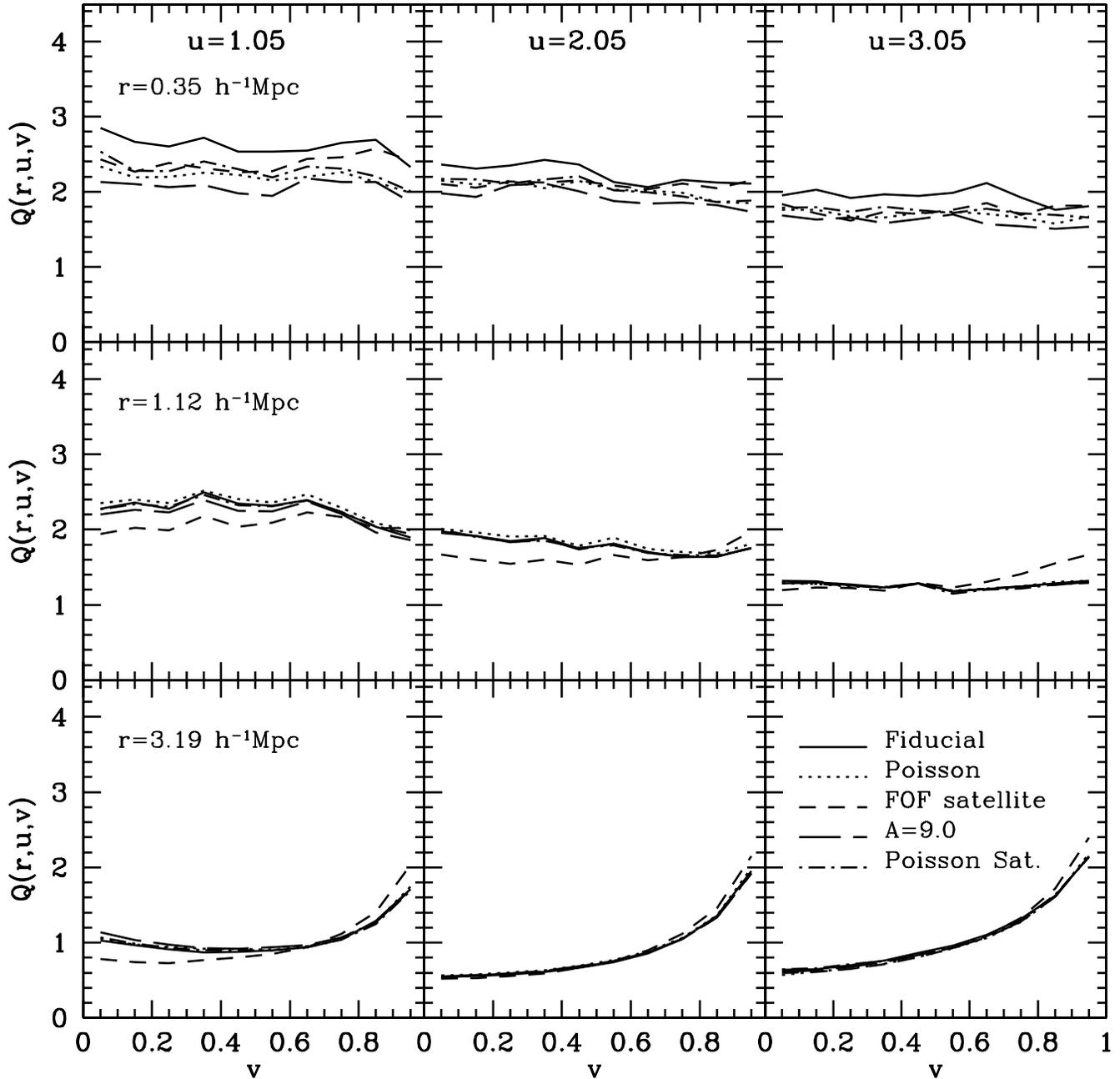,width=\hdsize}} 
\caption{The normalized 3PCFs for mock galaxies with
  various  assumptions about galaxy  distribution in  individual haloes
  (see  text  for  details).  The  line  styles  are  the same  as  in
  Fig.~\ref{fig:2pg-fof}.}
\label{fig:3pg-fof}
\end{figure*}

Next we  focus on the normalized  3PCF for galaxies.  As  for the dark
matter, we first consider  the equilateral triangle configuration. The
model    predictions    and    MGDs    results   are    compared    in
Fig.~\ref{fig:3pg-eq}. The solid and dashed lines are results obtained
by   assuming  $A=14$   and   $A=9$  for   satellite  galaxies   (with
$r_{\rm exc}=2.0\,r_{180}$ and  no truncation in the  halo mass function),
respectively.  The  dot-dashed curves  are model predictions  where an
upper limit  on halo  mass, $M= 2\times  10^{15} \msunh$ is  included. 
The open  squares with  errorbars are the  mean and  1-$\sigma$ errors
obtained  from  4  independent  MGDs.   As  one  can  see,  the  model
predictions  match the  simulation results  extremely  well. Comparing
these results with  the corresponding results for dark  matter, we see
that  the normalized  3PCFs  for  galaxies on  small  scales are  much
smaller  and with  a weaker  dependence on  $A$. And  unlike  for dark
matter,  the truncation  in the  halo mass  function does  not  have a
significant  impact on  the normalized  3PCF of  galaxies.   There are
several reasons for these results. Firstly, haloes that host less than
three galaxies  do not contribute  to the 1-halo  term in the  3PCF of
galaxies on small  scales, and so the 3PCF on small  scale is reduced. 
For the  same reason, the effect  of changing $A$  is reduced, because
the strong  dependence of $Q$ on $A$  for the mass on  small scales is
due to low-mass haloes. Finally, the 3PCF of galaxies is less sensitive
to the  truncation of the  mass function, because the  halo occupation
number  of galaxies in  massive haloes increases roughly  as $M^{0.8}$
(Yang \etal 2004).

Fig.~\ref{fig:3pg}  shows normalized  3PCFs, $Q(r,u,v)$,  for galaxies
with  absolute   magnitudes  $M_{b_J}  -  5\log  h<   -18.5$  and  for
non-equilateral triangle configurations.  Lines and symbols correspond
to model predictions and results obtained from the MGDs, respectively.
Overall the agreement is satisfactory. Compared to the   $Q(r,u,v)$ of 
dark matter particles, the  normalized 3PCF of galaxies, has a similar 
form, but with  systematically  lower amplitudes  on  small scales due 
to the reasons given above.

In Section  \ref{sec:2pcf-comp}, we compared the 2PCFs  for MGDs using
different models  for the spatial distribution of  satellite galaxies. 
Here,  we make a  similar comparison  for the  3PCF.  The  results are
shown in  Fig.~\ref{fig:3pg-fof}.  Compared with  the fiducial, $A=14$
model  (solid lines),  the $A=9$  model predicts  $Q(r,u,v)$  that are
lower  on  small scales  ($r<0.35\mpch$),  consistent  with the  model
predictions  shown in  the left-hand  panel of  Fig.~\ref{fig:3pg-eq}. 
The short-dashed  lines indicate  the results obtained  when assigning
satellite galaxies  the position of a randomly  selected particle from
the FOF group associated  with the halo under consideration. Comparing
these 3PCFs  on small  scales with those  from our fiducial  model, in
which we assume  spherical NFW distributions, gives an  idea as to how
sensitive the 3PCF is to non-sphericity of the spatial distribution of
galaxies  within  haloes.   Although   the  $Q(r,u,v)$  based  on  FOF
satellites are  slightly lower than  those of our fiducial  model, the
differences  are  small  ($\lta  20$  percent),  suggesting  that  the
assumption  of spherical  haloes  does not  lead  to large  systematic
errors. The dotted curve shows the normalized 3PCFs obtained
using Poisson sampling. Contrary to the 2PCF, the normalized 3PCF
is not very sensitive to the second-order moment of the halo 
occupation numbers.  Finally, the dot-dashed lines are the results
for the MGD in which the number of satellite galaxies in each halo 
is assumed to follow a Poisson distribution. The values of 
$Q(r,u,v)$ given by this model is quite similar to that 
given by the Poisson sampling model.
 
It is somewhat surprising that  $Q(r,u,v)$ is quite insensitive to all
these changes tested.   The reason might be that  these changes affect
the 2PCF and 3PCF in a  similar way, so that the effect is compensated
in the normalized  quantity, $Q(r,u,v)$. 

\subsubsection{Dependence on galaxy type and luminosity}
\label{sec:type}

Since  the CLF  models used  here contain  information  regarding both
galaxy type and luminosity, we can investigate how the 3PCF depends on
these quantities.

\begin{figure}
\centerline{\psfig{figure=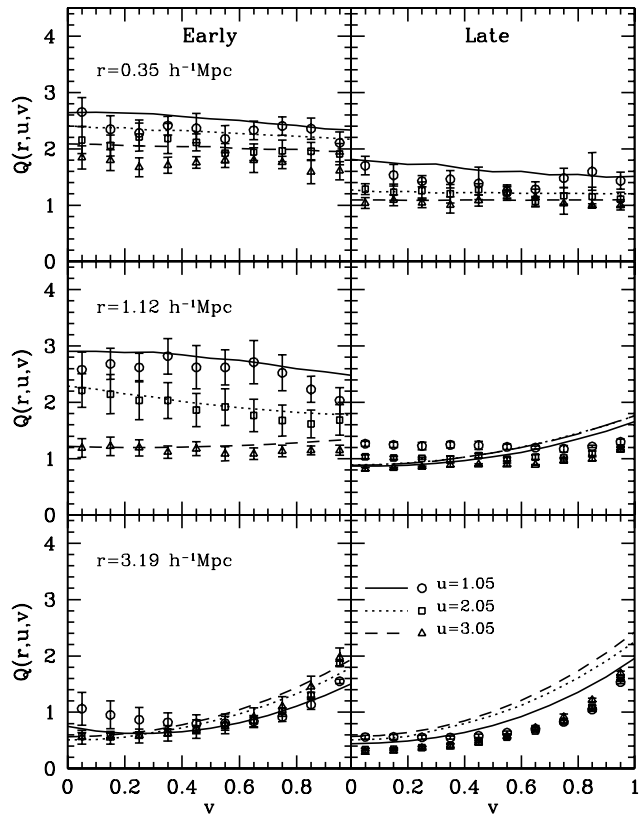,width=\hssize}}
\caption{The normalized 3PCFs for early- and late-type galaxies. }
\label{fig:3pg-type}
\end{figure}

Fig.~\ref{fig:3pg-type}  shows  the  $Q(r,u,v)$  obtained  for  early-
(left-hand  panels) and late-type  (right-hand panels)  galaxies using
both  the  halo  model  (lines)  and  the  MGDs  (symbols).   Overall,
$Q(r,u,v)$ for  early-type galaxies is systematically  higher than for
late-type  galaxies,   and  both  galaxy  types   reveal  a  different
dependence on the shape of  the triplet. The higher amplitude on small
scales for  the early-type galaxies reflects the  fact that early-type
galaxies in  our mock sample  are preferentially located in  clusters. 
The  3PCFs for  galaxies of  different  types have  been discussed  in
Takada \& Jain (2003), who  adopted the halo occupation models for red
and blue galaxies  of Scranton (2002).  Contrary to  our results, they
found that  red galaxies have  smaller $Q(r,u,v)$.  Note that  the CLF
model and  MGDs used  in our study  have been compared  carefully with
various  observations.   Therefore, we  are  confident  that our  halo
occupation models are more accurate.
 
\begin{figure}
\centerline{\psfig{figure=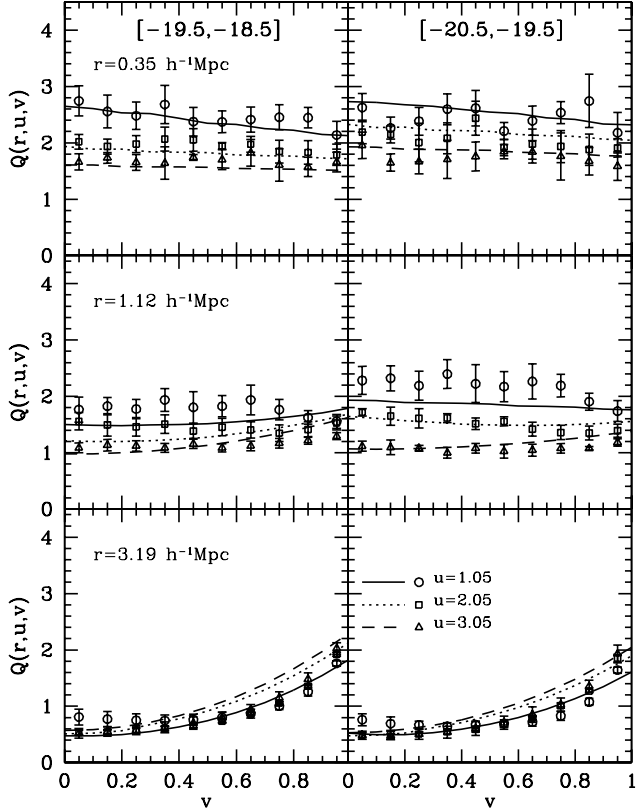,width=\hssize}}
\caption{The normalized 3PCFs for mock galaxies
  in two  absolute magnitude bins. The  range of $M_{b_J} -  5 \log h$
  covered by each bin is marked in the two upper panels. }
\label{fig:3pg-lum}
\end{figure}

We  also   investigate  the  luminosity   dependence  of  the   3PCF.  
Figure~\ref{fig:3pg-lum}  plots the normalized  3PCFs for  galaxies in
two different  luminosity ranges: $-19.5  <M_{b_J} - 5\log h  < -18.5$
and  $-20.5 <M_{b_J}-  5\log h  <  -19.5$.  Note  that the  luminosity
dependence of the  normalized 3PCFs is quite weak. This  is due to the
fact that  a large fraction  of relatively bright spiral  galaxies are
isolated  central galaxies of  galaxy-sized haloes. 

Unfortunately, we can  not test the luminosity dependence  of the 3PCF
for even  brighter galaxies.  For galaxies with  $M_{b_J} - 5\log  h <
-20.5$, for which the 2PCF  is much stronger than for fainter galaxies
(Yang  \etal 2004),  the number  of galaxies  is too  small to  give a
reliable estimate of the normalized 3PCF.

\subsubsection{The impact of the quadratic bias term}
\label{sec:quad}

 As mentioned in Section \S\ref{sec:haloclustering}, 
we have neglected the quadratic term of the halo bias relation.
Here we use simple considerations to assess the impact of  
this term on the predictions of the normalized 3PCFs. 
As a simple approximation, the normalized 3PCF of galaxies
on quasi-linear scales (where $\delta_{\rm mass}\ll 1$)
can be  related to that of dark matter as follows,
\begin{equation}
Q_{\rm g} = {1\over {\bar b}}\, Q_{\rm mass} \, 
+ {{\bar b_2} \over {\bar b}^2 }\,
\end{equation}
(Kayo \etal 2004). Here $\bar b$ is given by eq.~\ref{averbias} and 
\begin{equation}
\label{averbias2}
\overline{b}_2 = {1 \over \overline {n}_{\rm g}}
\int_{0}^{\infty} n(M) \, {\cal N}(M) \, b_2(M) \, {\rm d} M\,,
\end{equation}
with $b_2(M)$ being the quadratic term of the halo bias 
(Mo, Jing \& White 1997; Scoccimarro \etal 2001). 
Thus including the quadatic term on $Q_{\rm g}$
is to add a term ${\bar b_2} / {\bar b}^2 $ on large scales.
To get an idea how big this term is, we estimate the average 
of this quantity using our MGDs discussed in  this section. 
For the samples of all, early-type, late-type, bright, and faint 
galaxies discussed above, the values of ${\bar b_2} / {\bar b}^2 $ 
are $-0.16$, $-0.05$, $-0.29$, $-0.12$, and $-0.22$, respectively.
If this term is taken into account, the overall agreement 
between model prediction and simulation results may be   
improved slightly on large scales. 

\section{Comparison with observations}
\label{sec:obs}

\subsection{The 2dFGRS and mock galaxy redshift surveys}

We use the  final public data release from  the 2dFGRS, which contains
about  $250,000$  galaxies  with  redshifts  and  is  complete  to  an
extinction-corrected  apparent magnitude  of $b_J\approx  19.45$.  The
survey covers an area of  $\sim 1500$ square degrees selected from the
extended APM Survey (Maddox et al. 1996). The survey geometry consists
of two  separate declination strips  in the North Galactic  Pole (NGP)
and  the South Galactic  Pole (SGP),  respectively, together  with 100
2-degree fields  spread randomly in the southern  Galactic hemisphere. 
In this  paper,  we will use galaxies in the  NGP and SGP  to estimate 
the apparent-magnitude limit redshift space 3PCFs. Only those galaxies 
with redshift $0.01<z<0.2$, spectra quality $q \geq 3$, and   redshift 
completeness $>0.7$ are considered.   

In  order  to  carry  out   a  proper  comparison  between  model  and
observations,  we construct  mock galaxy  redshift  surveys (hereafter
MGRS) with the same selection  criteria and observational biases as in
the 2dFGRS.  We  follow the procedure used in Yang  et al.  (2004) and
stack various simulation boxes (of different sizes) together to sample
a  sufficiently large  volume  and with  sufficient resolution.   This
allows us to construct MGRSs with  the same depth (in redshift) as the
2dFGRS  and with  full sampling  of  the luminosity  function down  to
$M_{b_J} - 5  \log h = -13.5$. Observational  selection effects in the
2dFGRS,  such as  position  dependent magnitude  limits, position  and
magnitude  dependent completeness, errors  in magnitude  and redshift,
are all  taken into  account (see  Yang \etal 2004  and van  den Bosch
\etal 2004 for details).  Using  the full set of numerical simulations
available to us, we construct eight independent MGRSs which we use for
comparison with the 2dFGRS.

\subsection{The redshift-space 3PCF of galaxies}

We  calculate the  3PCFs in  redshift-space using  the same  method as
described  in  Section   \ref{sec:3PCF1},  except  that  redshifts  of
galaxies are used as distances  in calculating the number of triplets. 
To estimate the number of  triangles for the random distribution, i.e. 
in estimating $RRR$, we use random  samples that are 32 times as dense
as the observational sample for  $s\le 2.0\mpch$  (here $s$ is
the length  of the shortest  side of the  triplet in redshift  space). 
For $2.0<s\le 5.0\mpch$, the random sample is 8 times as dense as
the  observational sample,  while  for $s>5.0\mpch$,  the
random samples have the same density as the observational sample.

We  use   the  same  method  as  in   equations  (\ref{eq:3pcf2})  and
(\ref{eq:Qruv})  to  calculate  the  normalized  redshift-space  3PCF,
$Q(s,u,v)$.   In  order to  take  account  of observational  selection
effects,        each       triangle        is        weighted       by
$[w_1(s_{12})w_2(s_{23})w_3(s_{31})
+w_1(s_{13})w_3(s_{32})w_2(s_{21})]/2$, where
\begin{equation}
w_i(s_{ij})=\frac{1}{1+4\pi n(z_i)J_3(s_{ij})}\,.
\end{equation}
Here $n(z)$ is  the density of galaxies as a  function of redshift and
$J_3(s)=\int_0^{s}  \xi(x) x^2  d x$,  where we  follow  Hawkins \etal
(2003) by adopting $\xi(s)={(s/13)}^{-0.75}$. 
 
The overall normalization in equation (\ref{eq:3pcf2}), i.e. the ratio
between $N_R$  and $N_D$, is  estimated by summing over  random points
and  galaxies,  each  weighted   by  $1/n(z_i)$.   For  the  two-point
correlation  function used  in  $Q(s,  u, v)$,  we  use the  estimator
proposed by Hamilton (1993),
\begin{equation}
\xi(s)=\frac{DD\times RR}{DR^2}-1\,,
\end{equation}
where $DD$ is the sum of galaxy-galaxy pairs with separation $s$, $RR$
and $DR$  are the sums  of random-random and galaxy-random  pairs with
the  same   separation,  respectively.   Each  pair   is  weighted  by
$w_i(s_{ij})w_j(s_{ij})$.

\subsection{Results}

\begin{figure*}
\centerline{\psfig{figure=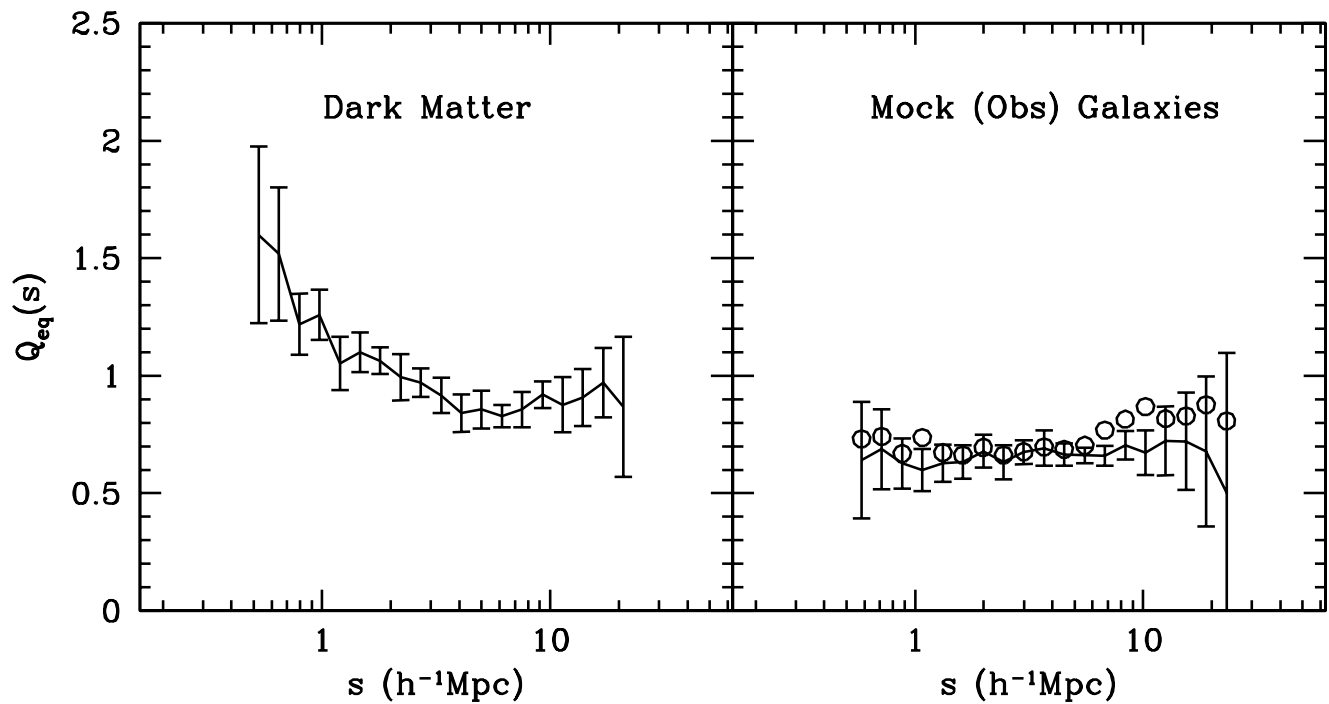,width=\hdsize}} 
\caption{
  The redshift-space 3PCFs of equilateral triangles for 2dFGRS 
  galaxies (circles in the right panel) compared with model 
  predictions. The left panel shows the  results for  mock 2dFGRS 
  samples  which use  dark matter  particles as galaxies.  The lines 
  with errorbars  in the right panel are results  for  mock  2dFGRS 
  samples  based  on  the CLF  model. The errorbars  are
  1-$\sigma$ variance among 8 independent mock samples.}
\label{fig:3pcf2dF_eq}
\end{figure*}
\begin{figure}
\centerline{\psfig{figure=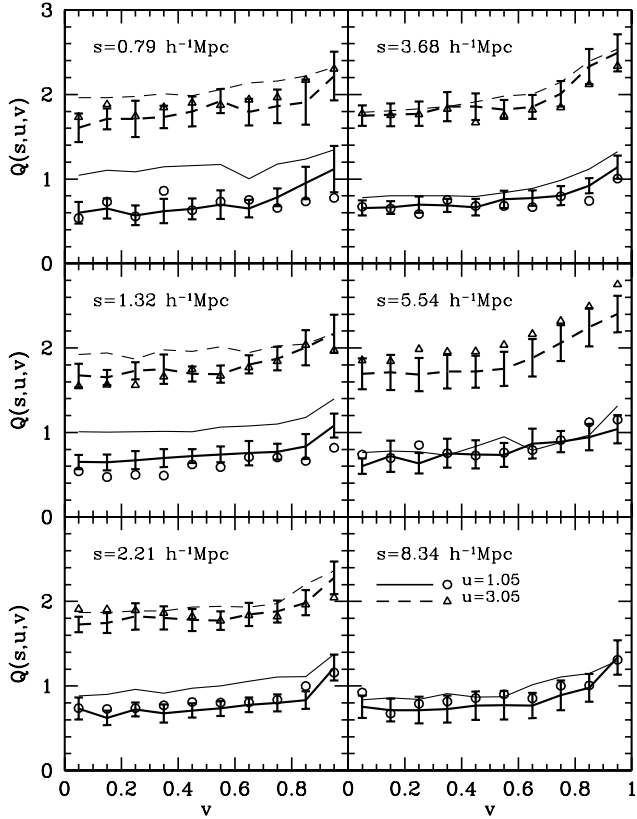,width=\hssize}} 
\caption{
  The redshift-space 3PCFs for 2dFGRS galaxies (symbols) compared with
  model predictions.   The thick lines with errorbars  are results for
  mock  2dFGRS samples  based  on  the CLF  model.  The errorbars  are
  1-$\sigma$ variance among 8 independent mock samples. The thin lines
  are  results for  mock samples  which use  dark matter  particles as
  galaxies.  For clarity,  the results for $u=3.05$ are  shifted up by
  $1.0$.}
\label{fig:3pcf2dF}
\end{figure}

We first consider the normalized 3PCFs of equilateral 
triangles, $Q_{\rm eq}(s)$. The 2dFGRS results are shown as the open
circles in the right panel of Fig.~\ref{fig:3pcf2dF_eq}. The lines 
with errorbars in the right panel are results
for the mock samples, with the errorbars giving the 1-$\sigma$ scatter
among the 8 MGRSs. Over all scales, the redshift-space 3PCFs 
obtained from our MGRSs  are in  good agreement with the observational 
results. We emphasize that this is a success of our CLF model   
in accounting  the bias of galaxy distribution relative 
to the mass. To demonstrate this more clearly, we  construct  
MGRSs in  which  dark matter  particles are randomly chosen to 
represent the `galaxy population', with each particle assigned 
a luminosity according to the   2dFGRS  luminosity function.   
Thus, the `galaxy distribution' is un-biased in these MGRSs.
The lines with errorbars in the left panel of Fig.~\ref{fig:3pcf2dF_eq} 
shows the normalized, redshift-space  3PCF thus obtained. 
This 3PCF is very different from that given by the mock sample 
based on the CLF, which reflects the effect of the bias in the 
distribution of galaxies relative to the mass.

Symbols in Fig.~\ref{fig:3pcf2dF} shows  the normalized 3PCF, $Q(s, u,
v)$, as a function of $s$, $u$ and $v$ obtained from the 2dFGRS.  Jing
\& B\"orner (2004) recently estimated  the 3PCFs of galaxies using the
2dFGRS early  data release, and  their results for  the redshift-space
3PCFs are similar to ours.  The thick lines with errorbars are results
for the mock samples based on CLF model. Again, the  redshift space 
3PCFs  $Q(s, u, v)$ of our MGRSs  are in  good agreement
with the 2dFGRS observational results on all scales and for different
triangle configurations. The thin lines in Fig.~\ref{fig:3pcf2dF} 
correspond to MGRSs using dark matter particles as galaxy tracers. 
Consistent with the results shown in Fig.~\ref{fig:3pcf2dF_eq}, 
these MGRSs predict 3PCFs that are in poor agreement with the 
2dFGRS observations, especially on small scales 
(see also Jing \& B\"orner 1998, 2004). 

\begin{figure}
\centerline{\psfig{figure=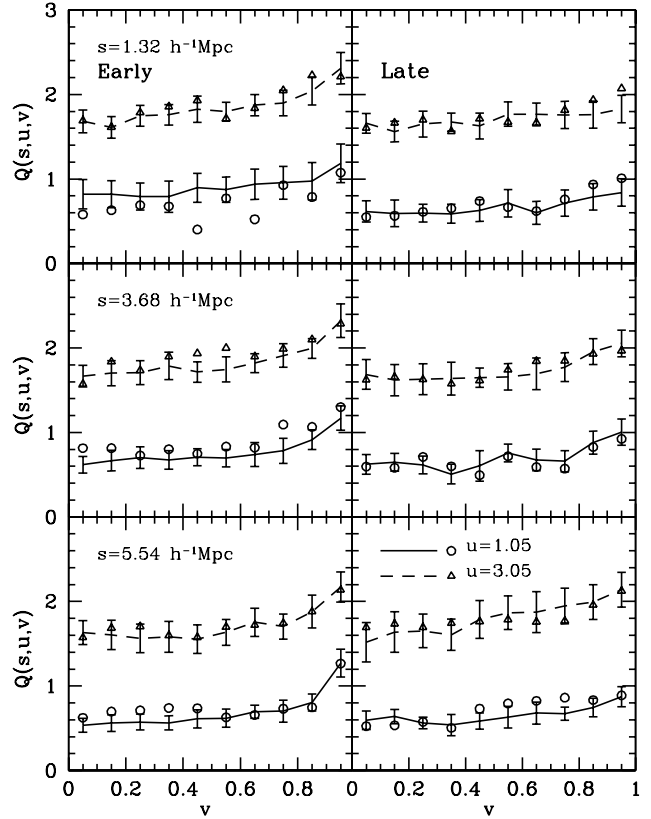,width=\hssize}}   
\caption{   The
  redshift  space 3PCFs  for  early-type (left  panels) and  late-type
  (right panels) galaxies.  The  symbols are results obtained from the
  2dFGRS.   The  lines with  errorbars  are  results  for mock  2dFGRS
  samples  based  on the  CLF  model.   The  errorbars are  1-$\sigma$
  variance among 8 independent mock samples.  For clarity, the results
  for $u=3.05$ are shifted up by $1$.}
\label{fig:3pcftype}
\end{figure}

\begin{figure}
\centerline{\psfig{figure=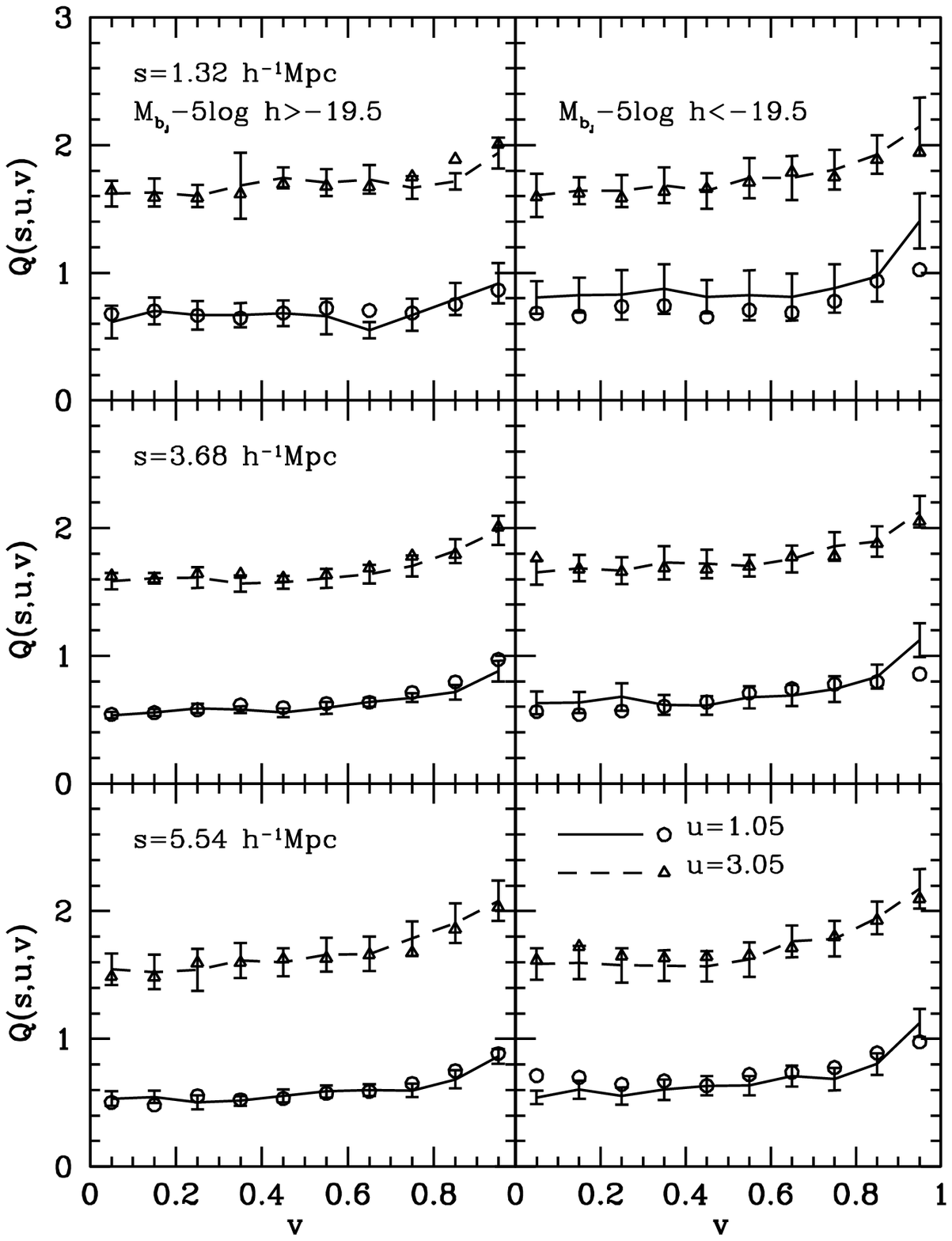,width=\hssize}} 
\caption{
  The redshift space  3PCFs for faint (left panels)  and bright (right
  panels) galaxies. Note that for the faint sample, only galaxies with
  redshifts $0.01<z<0.12$ are used.   Symbols are the results obtained
  from the 2dFGRS.   Lines with errorbars are results  from the 2dFGRS
  mock samples  based on the  CLF model. The errorbars  are 1-$\sigma$
  variance among 8 independent mock samples.  For clarity, the results
  for $u=3.05$ are shifted up by $1$.}
 \label{fig:3pcflum}
\end{figure}

Fig.~\ref{fig:3pcftype}  shows $Q(s,u,v)$  for  galaxies of  different
types.   Note  that  early-type  galaxies have  only  slightly  higher
$Q(s,u,v)$ than  late-type galaxies.  The  strong type-dependence seen
in the real-space  3PCF is not seen here,  mainly because the velocity
dispersion  of galaxies  in individual  clusters tends  to  reduce the
correlations on small scales. Once again, the $Q(s,u,v)$ obtained from
our MGRSs are  in excellent agreement with the  observations, for both
early- and  late-type galaxies. Finally,  Fig.~\ref{fig:3pcflum} shows
that there is no significant  luminosity dependence of $Q(s, u, v)$ in
the  data, a  result  that, once  more,  matches well  with our  model
prediction. Note that these  type and  luminosity dependences are also 
found in various recent works (Jing \& B\"orner 2004; Kayo et al. 2004).

\section{Conclusions}
\label{sec:concl}

In  this  paper,  we  have  used  the halo  model  combined  with  the
conditional  luminosity function  formalism to  predict the  3PCFs for
both dark matter and galaxies. These analytical model predictions have
been  compared  with results  obtained  from high-resolution  $N$-body
simulations and from mock  galaxy distributions constructed from these
simulations.  With  proper assumptions, the  halo model can  match the
mass  3PCF reasonably  well.  On  small, non-linear  scales  ($\la 0.5
\mpch$)  the 3PCF is  contributed mainly  by the  1-halo term,  and is
sensitive  to  the  concentration  of  the  dark  matter  haloes.   On
intermediate scales ($r\sim  2 \mpch$) where both the  1-halo term and
2-halo  term  contribute  significantly,  the  3PCF  is  sensitive  to
abundance of  the few  most massive  haloes. Due to  the way  in which
galaxies  are  biased  with  respect  to  the mass,  and  due  to  the
discreteness in the galaxy  distribution, both effects are much weaker
for galaxies  than for mass.  Overall, galaxies  have lower normalized
3PCFs than dark  matter, which is mainly due  to a strongly suppressed
1-halo term.

We  also investigate the  dependence of  the 3PCF  on galaxy  type and
luminosity.   Since  early-type   galaxies  reside  preferentially  in
massive  haloes,  they have  higher  normalized  real-space 3PCF  than
late-type galaxies.  The dependence  on galaxy luminosity, however, is
found to be much weaker. 

Finally,  we  have  compared   the  redshift-space  3PCF  of  galaxies
predicted by current CDM model with the observational results obtained
from  the 2dFGRS. We  have shows  that, with  the more  realistic mock
samples based on the CLF in  dark haloes, the model predictions are in
good  agreement with  observations in  redshift space.   These results
provide further  support to  the CLF  model advocated by  Yang et  al. 
(2003) and van den Bosch et al. (2003).


\section*{Acknowledgement}

We thank  Y.P.  Jing  \& Y. Suto for useful comments and discussions.   
XY is supported  by NSFC
(No.10243005) and by USTCQN.  Numerical simulations used in this paper
were  kindly  provided by  Y.P.  Jing, and  were  carried  out at  the
Astronomical Data Analysis Center  (ADAC) of the National Astronomical
Observatory, Japan.


\end{document}